\documentclass[twocolumn,showpacs,preprintnumbers,amsmath,amssymb]{revtex4}
\usepackage{amssymb,amsmath}
\usepackage{graphics}
\usepackage{epstopdf}
\usepackage{epsfig}
\usepackage{color}
\usepackage{amsfonts}

\usepackage{amscd}

\begin{document}

\title{ \bf A non-Markovian decoherence theory for double dot charge qubit}

\author{\bf Matisse W. Y. Tu and Wei-Min Zhang}
\email{wzhang@mail.ncku.edu.tw} \affiliation{Department of Physics
and Center for Quantum Information Science, National Cheng Kung
University, Tainan 70101, Taiwan } \affiliation{National Center
for Theoretical Science, Tainan 70101, Taiwan}

\affiliation{}

\date{September 20, 2008}

\begin{abstract}
In this paper, we develop a non-perturbation theory for describing
decoherence dynamics of electron charges in a double quantum dot
gated by electrodes. We extend the Feynman-Vernon influence
functional theory to fermionic environments and derive an exact
master equation for the reduced density matrix of electrons in the
double dot for a general spectral density at arbitrary temperature
and bias. We then investigate the decoherence dynamics of the
double dot charge qubit with back-action of the reservoirs being
fully taken into account.  Time-dependent fluctuations and leakage
effects induced from the dot-reservoir coupling are explicitly
explored. The charge qubit dynamics from the Markovian to
non-Markovian regime is systematically studied under various
manipulating conditions. The decay behavior of charge qubit
coherence and the corresponding relaxation time $T_1$ and
dephasing time $T_2$ are analyzed in details.

\end{abstract}


\maketitle

\section{Introduction}
Double quantum dot systems have been attracting much attention
because of their intriguing properties and their potential
applications in nanotechnology and quantum information processing
\cite{prs,rmp}. The basic structure of a double quantum dot system
can be viewed as electrons confined in an electrostatic potential of
double wells created by the fabricated gates, source and drain
electrodes in the heterostructure of a semiconductor. The
heterostructure of GaAs/AlGaAs is a typical example for the
realization of gate-defined double dots.  The tunability of various
couplings and energy levels in the dots makes it a promising a
quantum device (see, for examples,
\cite{Elzerman,hayashi,petta,Ludwig,Gorman}). Maintaining electron
coherence in a double quantum dot is an important ingredient in
making it part of a quantum information processor. However,
fluctuations and dissipations brought up by quantum operations of
manipulations and measurements as well as various features of the
material enrich the physics of double dot systems more than perfect
coherent evolution. Thus a lot of attentions have been paid to
investigate how various noises and interactions with the
surroundings attenuate the coherent evolution of electrons in the
double dot.

In this paper, we will concentrate on a non-perturbative dynamical
theory for charge qubit manipulation with a double dot system
gated by electrodes. In the quantum computing scheme in terms of
double dots where the electron charge degree of freedom is
exploited, the effects in deviating the coherency of charge
dynamics are summarized in the fluctuations of the inter-dot
coupling and energy splitting between the two local charge states
as well as the dissipation induced damping effects. The amplitudes
of these fluctuations can be estimated from measurements of the
noise spectrum of electron currents and the minimum line width of
elastic current peak \cite{prs}. Parallel theoretical works have
been developed with different approaches in the literature for the
purposes of both simulating the experimental results and
understanding the physical mechanisms living in the double dot. In
the present work, we shall extend the Feynman-Vernon influence
functional theory \cite{fv63} to fermionic environments and derive
an exact master equation describing the coherent and decoherent
dynamics of the electron charges in the double dot with the
back-action effects of the reservoirs being fully taken into
account.

Stochastic noise processes resulted in a time dependent
Hamiltonian for the double dot have been widely analyzed in
simulating the charge dynamics under noise influences. The
Bloch-type rate equations for describing the double dot transport
properties have been investigated by Gurvitz and co-worker
\cite{Gurvitz} used the many-body Schr$\ddot{\mbox{o}}$dinger
equation approach which is further applied to study decoherence of
double dot charge qubit in \cite{Fujisawa04}. The phonon assisted
processes have been investigated within the Born-Markov regime by
Brandes {\it et al.} using Born-Markov typed master equation
\cite{tb}. A general expression of the qubit density matrix in
case of pure dephasing \cite{Palma} was used by Fedichkin and
Fedorov \cite{Fedorov} to study the error rate of the charge
qubit. Stavrou and Hu \cite{Stavrou} considered in details the
wavefunctions of the double dot charge qubit for decoherence
analysis. Karrasch, {\it et al.} come with the functional
renormalization group approach in dealing with the transport
aspects of the multiple coupled dots \cite{ck}. The non-Markovian
dynamics has also recently been studied by a suitable spin-boson
model considering the acoustic phonons by Thorwart, {\it et al.}
using numerical quasiadiabatic propagator path integral scheme
\cite{mt,Liang}. Without Born-Markov approximation, Wu {\it et
al.} had devised an analytical expression for the dynamical
tunneling current using a perturbation treatment based on a
unitary transformation \cite{Wu,xf}.  Effects from Coulomb
interaction between the dots and the gate electrodes with the
formulation of kinetic equations have been presented by Woodford
{\it et al.} \cite{sr}. The diversity in the methodologies and
issues concerned in the literature show the physical richness of
this novel system.

\begin{figure}[ht]
%\begin{center}
\begin{center}
\includegraphics[width=4.5cm, height=2.5cm]{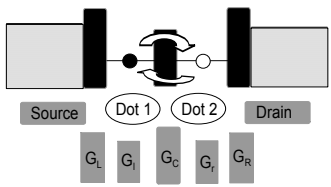}
\caption{A schematic plot for the double quantum dot system}
\label{fig1}
\end{center}
%\end{center}
\end{figure}

As we can see there are many factors competing to play the
consequent physics in the double quantum dot system.  To single
out one factor from the others on the resulted dynamical
properties of this charge device, we shall first concentrate in
this paper on the effects induced by dot-reservoir coupling, where
the double dot is designed in the strong Coulomb blockade regime
such that each dot only contains one energy level. The reservoirs
consist of the source and drain electrodes which are controllable
through the bias voltage. A schematic plot of the system is shown
in Fig.~\ref{fig1}.
Thus the total Hamiltonian of the system we concern in this work
is given by
\begin{align}
\label{hamil0} H & = E_1 a^\dagger_1 a_1 + E_2 a^\dagger_2
a_2+T_c( a_{2}^{\dagger}a_{1}+a_{1}^{\dagger}a_{2})\nonumber
\\ & ~~~~~ +
\sum_{k} (\varepsilon_{L k}a^{\dagger}_{L k}a_{L k} +
\varepsilon_{R k}a^{\dagger}_{R
k}a_{R k} ) \nonumber \\
& ~~~~~+ \sum_{k}(t_{1Lk}a^{\dag}_{1}a_{Lk} +t_{2Rk} a^{\dag}_{R
k}a_{2} + {\rm H.c})
\end{align}
which contains the Hamiltonians of the double dot, the source and
drain electrodes plus the interaction (electron tunneling processes)
between them. The notations follow the convention and will be
specified in details later.

Our treatment is based on the exact master equation we derived for a
general spectral density of the electron reservoirs at arbitrary
temperature and bias:
\begin{align}
\dot{\rho}=&-i[H'(t),\rho]
+\Gamma^0(t)\rho \nonumber \\
& + \sum_{ij}\Gamma_{ij}(t)(2a_{j}\rho a^{\dag}_{i}-
a^{\dag}_{i}a_{j}\rho-\rho a^{\dag}_{i}a_{j}) \nonumber\\&
+\sum_{ij}\Gamma^\beta_{ij}(t)(a_{j}\rho a^{\dag}_{i} -
a^{\dag}_{i}\rho a_{j}- a^{\dag}_{i}a_{j}\rho-\rho
a^{\dag}_{i}a_{j}) , \label{m-e}
\end{align}
where $\rho$ is the reduced density matrix of the double dot
obtained from the full density matrix of the double dot plus the
reservoirs by tracing out the environmental degrees of freedom, and
\begin{align}
H'(t)&=E'_{1}(t)a^{\dag}_{1}a_{1}+E'_{2}(t) a^{\dag}_{2}a_{2}+
T'_c(t) a^{\dag}_{1}a_{2}+T'^*_c(t)a^{\dag}_{2}a_{1}
\end{align}
is the corresponding effective Hamiltonian. All the time-dependent
coefficients in the above equations will be derived explicitly and
non-perturbatively in Sec.~III. The time-dependent fluctuations of
the energy levels, $E'_i(t)$, and the inter-dot transition
amplitude, $T'_c(t)$, are the renormalization effects risen from
the electron tunneling processes between the double dot and the
reservoirs. Other non-unitary terms describe the dissipative and
noise processes with time-dependent coefficients, $\Gamma^0(t),
\Gamma(t)$ and $\Gamma^\beta(t)$, depicting the full non-Markovian
decoherence dynamics. Eq.~(\ref{m-e}) is obtained without
considering the inter-dot Coulomb repulsion. But as we will show
explicitly in Sec. III it is easy to extend to the strong
inter-dot Coulomb repulsion regime where the strong inter-dot
Coulomb repulsion simply leads one to exclude the states
corresponding to a simultaneous occupation of two dots from
Eq.~(\ref{m-e}). Then the exact master equation allows us to
exploit the intrinsic quantum decoherence effects in the electron
charge coherency brought up by the tunneling processes between the
dots and the reservoirs through the bias controls.

The master equation (\ref{m-e}) is derived by extending the
Feynman-Vernon influence functional theory \cite{fv63} to fermion
coherent state path integrals \cite{wmz90}. Historically, since it
was first developed by Feynman and Vernon in 1963 for quantum
Brownian motion (QBM) modelled as a central harmonic oscillator
linearly coupled to a set of harmonic oscillators simulating the
thermal bath, the influence functional theory has been widely used
to study dissipation dynamics in quantum tunneling problems
\cite{legt} and decoherence problems in quantum measurement theory
\cite{zuk1,zuk}. In these early applications, the master equation
was derived for some particular class of ohmic environment
\cite{cal,hak,uz89}. The exact master equation for the QBM with a
general spectral density at arbitrary temperature was obtained by
Hu and co-workers in 1992 \cite{hu}. Applications of the QBM exact
master equation cover various topics, such as quantum decoherence,
quantum-to-classical transition and quantum measurement theory,
etc.\cite{zuk}. Very recently, such an exact master equation is
further extended to the system of two entangled optical fields and
two entangled harmonic oscillators for the study of non-Markovian
entanglement dynamics in quantum information processing
\cite{an07,chou08}. Nevertheless, using the influence functional
theory to derive the exact master equation has been largely
focused on the bosonic type of environments, up to date.

On the other hands, the development of quantum transport theory in
nanosystems has continuously attracted attention in the last two
decades because of the great achievements in nanotechnology, where
the reservoir is, in many cases, a fermion system. The traditional
approach to study the quantum transport in nanosystems is
Schwinger-Keldysh's nonequilibrium Green function formulism
\cite{schwinger,keldysh} which has been extensively used in
successfully describing various quantum transport phenomena, such
as Kondo effect, Fano resonance and Coulomb blockade effects in
quantum dots \cite{hers91,lee93,konig96,fuji03}. Master equations
for quantum transport through quantum dots have also been derived,
but mostly in the perturbation theory up to the second order
\cite{brud91,leh02,ped05,li05,har06}. The exact master equation
can be, in principle, obtained through the real-time diagrammatic
expansion approach developed by Sch\"{o}n and co-workers
\cite{schon94,konig96}, as shown recently by one of the authors
and co-worker \cite{lee07}. Another interesting formulism is the
recently published hierarchical expansion of the equations of
motion for the reduced density matrix by Yan {\it et. al.}
\cite{yan08}. Nevertheless, in contrast to the bosonic
environments \cite{hu}, an explicit formula of the exact master
equation for fermionic environments with a general spectral
density at arbitrary temperature and bias has not been carried out
except for Eq.~(\ref{m-e}) in this work.

Unlike the quantum transport in nanosystems where people pay more
attention on the tunneling current spectrum and its statistics,
one cares in quantum information processing how the qubit
coherency can be maintained for fast quantum operations where a
strong coupling is required. Then a nonperturbative (with respect
to the coupling between the system and its environment) master
equation is more desirable for the precision manipulations of
qubit states. Eq.~(\ref{m-e}) obtained in this paper has fully
taken into account the back-action of the electron reservoirs at
arbitrary temperature and bias. This master equation is also valid
for a general spectral density. It is fully non-perturbative that
goes far beyond the Born-Markovian approximation often used in the
literature. It enables us to explore the dynamics of the electron
charge coherence in the double dot, from Markovian to
non-Markovian regime under various manipulating conditions. Many
other approximated master equations that have been developed in
the literature can be obtained at well defined limits of the
present theory.

The rest of the paper is organized as follows.  In the next section,
we will use fermion-coherent-state path integral approach to solve
exactly the electron dynamics in an isolated double quantum dots, as
an illustration. We then extend the Feynman-Vernon influence
functional theory originally built on the coordinate representation
in quantum mechanics to fermion coherent state representation. The
exact master equation for the reduced density matrix of the double
dot system coupling to electron reservoirs is derived in Sec. III,
where we also reproduce many other approximate master equations at
well defined limits of the present formulae. In Sec. IV, we
investigate the non-Markovian decoherence dynamics of this device
including the tunneling induced fluctuations in the energy splitting
and inter-dot coupling of the double dot as well as the noise and
dissipation effects on the charge mode populations and
interferences. The leakage effect is also discussed together there.
The decay behaviors of charge qubit coherence and the corresponding
relaxation time $T_1$ and dephasing time $T_2$ are analyzed in
details. Conclusive remarks are given in Sec. V, and Appendices are
presented for some detailed derivations.

\section{Fermion Coherent State Path Integral Approach to
an isolated Double Dot} To illustrate the fermion coherent state
path integral approach to the electron dynamics in a double
quantum dot, we shall consider in this section a simple solvable
system, a single electron in an isolated double dot, before we go
to explore the realistic system in the next section. We also
assume that each of the dots contains only one energy level, $E_1$
and $E_2$ respectively.
 The Hamiltonian of this isolated double quantum dot is
\begin{equation}
H=E_1a^{\dag}_1a_1+E_2a^{\dag}_2a_2+T_c(a_2^{\dagger}a_1+a_1^{\dagger}a_2).
\label{isoh}
\end{equation}
The notations follow the convention: $a^\dagger_{1,2}$ are the
creation operators for electrons in the double dot, and $T_c$ is
the electron transition amplitude between the dots. In terms of
the density operator, the time evolution of the system is
described by
\begin{equation}
\rho(t)=U(t-t_0)\rho(t_0)U^\dagger(t-t_0) ,
\end{equation}
where the density matrix $\rho(t)$ is the state of the system at a
later time $t$, $\rho(t_0)$ is the state at the initial time
$t_{0}$, $U(t-t_0)=\exp{\big\{-{i\over \hbar}H(t-t_0)}\big\}$ is
the evolution operator of the system, and we let $\hbar=1$
hereafter.

Using the fermion coherent state representation \cite{Faddeev},
the density matrix at time $t$ is expressed as
$\langle\xi|\rho(t)|\eta\rangle=\rho(\xi^*,\eta,t)$ where $\xi$
and $\eta$ are two Grassmann variables characterizing the two-mode
fermion coherent states,
\begin{equation}
|\xi\rangle = \prod_{i=1,2} |\xi_i\rangle~,~~ |\xi_i\rangle = \exp
( -\xi_i a^\dagger_i ) | 0 \rangle ,
\end{equation}
The fermion coherent state defined above is an eigenstate of the
fermion annihilation operator, $a_{i}|\xi_{i}\rangle
=\xi_{i}|\xi_{i}\rangle $. As these coherent states are
over-complete, they obey the resolution of identity, $\int d\mu
(\xi)|\xi \rangle \langle \xi |=1,$ where the integration measure is
defined by $d\mu (\xi)=\prod_i e^{-\xi^*_{i} \xi_{i}} d\xi^*_i
d\xi_{i}$. Note that the fermionic coherent states we used here are
not normalized, and the normalization factors are moved into the
above integration measure. Moreover, these coherent states are also
nonorthogonal. The overlap of two fermionic coherent states is:
$\langle \xi|\xi'\rangle = \exp (\xi^\dagger\xi')$ with a matrix
notation $\xi^{\dagger}=(\xi^{*}_{1}~\xi^{*}_{2})$.

The use of the coherent-state representation makes the path
integral formulation for a fermion system generally possible. In
the fermionic coherent-state representation, the time evolution of
the density matrix becomes
\begin{eqnarray}
\rho(\xi_f^*,\eta_f,t) =\int d\mu(\xi_0)d\mu(\eta_0)
\rho(\xi_0^*,\eta_0,t_0) \nonumber \\ \times
J(\xi_f^*,\eta_f,t|\xi_i0,\eta^*_0,t_0) ,
\end{eqnarray}
where $J(\xi_f^*,\eta_f,t|\xi_0,\eta^*_0,t_0)=
\langle\xi_f|U(t-t_0)|\xi_0\rangle \langle\eta_0|
U^\dagger(t-t_0)|\eta_f\rangle $ is the propagating function in
which the forward and backward transition amplitudes
$\langle\xi_f|U(t-t_0)|\xi_0\rangle$ and $\langle\eta_0| U^\dagger
(t-t_0)|\eta_f\rangle$ can be solved exactly using the path
integral for the Hamiltonian (\ref{isoh}).

Explicitly, the fermion coherent state path integral for the
forward transition amplitude is given by
\begin{equation}
\langle\xi_f|U(t-t_0)|\xi_0\rangle= \int\mathcal{D}[\xi,\xi^*]\exp
\big(i S_c[\xi^*,\xi] \big), \label{trsiamp}
\end{equation}
where the action is
\begin{align}
& S_c[\xi^*,\xi] = \sum_{i=1,2} \Big\{- {i \over 2} \big[\xi^*_{i
f}\xi_{i}(t)+\xi^*_{i}(t_0)\xi_{i 0} \big] ~~~~~~~~~
\nonumber \\
& ~~~ +\int_{t_0}^{t}d\tau \big[ {i \over 2} (
\xi^*_{i}\dot{\xi}_{i} - \dot{\xi}_{i}^*\xi_{i}) -(E_i
\xi^*_i\xi_i+T_c \xi^*_i\xi_{i'}) \big]\Big\} . \label{action}
\end{align}
In the above equation, the path integral $\mathcal{D}[\xi,\xi^*]$
integrates over all pathes $\xi_i(\tau)$ and $\xi^*_i(\tau)$
bounded by $\xi_i(t_0)=\xi_{i0}$ and $\xi^*_i(t)=\xi^*_{i f}$ with
$i \ne i'$. Since the action in Eq.~(\ref{trsiamp}) has a
quadratic form, the path integral can be exactly carried out with
the stationary path method \cite{fey65,Faddeev}. The result is
\begin{align}
\langle\xi_f|U(t-t_0)|\xi_0\rangle=
 \exp{\sum_i \Big[
{\xi^*_{i f}\xi_{i}(t)+\xi_{i}^*(t_0)\xi_{i 0}\over2}
 \Big]},
\end{align}
where $\xi_i^*(t_0)$ and $\xi_i(t)$ are determined by the solution
of the equations of motion
\begin{subequations}
\label{EM-0}
\begin{align}
\dot{\xi}_{i}(\tau) +
i\big[E_{i}\xi_{i}(\tau)+T_c\xi_{i'}(\tau)\big]&=0,
\label{EM-0a} \\
\dot{\xi}^*_{i}(\tau)-i\big[E_{i}\xi^*_{i}(\tau)
+T_c\xi^*_{i'}(\tau)\big]&=0, \label{EM-0b}
\end{align}
\end{subequations}
with the boundary conditions $\xi_{i}(t_0)=\xi_{i 0}$ and
$\xi^*_{i}(t)=\xi^*_{i f}$, where $i \neq i'$. The backward
transition amplitude $\langle\eta_0|
U^\dagger(t-t_0)|\eta_f\rangle$ can be found by the same
procedure. From Eq.~(\ref{EM-0}) we can introduce a $2\times 2$
matrix $u_0(t)$ such that $\xi(t)=u_0(t)\xi_{0}$ and
$\xi^{\dagger}(t_0)=\xi^{\dagger}_{f}u_0(t)$ where $u_0(t)$
satisfies the equation of motion
\begin{align}
\dot{u_0}(t)+i\begin{pmatrix}E_{1}&T_c\\
T_c&E_{2}\end{pmatrix}u_0(t)=0 \label{fsx}
\end{align}
with the boundary condition $(u_0)_{ij}(t_0)=\delta_{ij}$. The
solution of (\ref{fsx}) is:
\begin{align}
u_0(t)= e^{-i\phi(t)}(\cos\varphi(t) I - i {\bf n}\cdot
\boldsymbol{\sigma} \sin\varphi(t))
\end{align} with
$\phi(t)=E(t-t_0)$, $\varphi(t)=\Omega_0(t-t_0)/2$. Here we have
also defined $E\equiv {1\over 2}(E_{1}+E_{2})$ and the Rabi
frequency $\Omega_0=\sqrt{\varepsilon^2 + \Delta^2}$, where
$\varepsilon \equiv E_{1}-E_{2}$ is the energy level splitting in
the double dot and $\Delta=2T_c$ the inter-dot tunnel coupling.
$\boldsymbol{\sigma}$ is the Pauli matrix, $I$ a $2 \times 2$
identity matrix and the unit vector ${\bf n} \equiv (\Delta,
0,\varepsilon)/\Omega_0$. Then the propagating function becomes
\begin{align}  \label{ppg0s}
J(\xi_{f}^*,\eta_{f},t|\xi_0,\eta^*_0,t_0)=
\exp{(\xi^{\dag}_{f}u_0(t)\xi_{0}+\eta^{\dag}_{0}u_0^{\dag}(t)\eta_{f})}.
\end{align}

If we use further the D-algebra of fermion creation and
annihilation operators in the fermion coherent state
representation $|\xi\rangle$ \cite{wmz90},
\begin{subequations}
\label{Dal}
\begin{align}
& \xi_i |\xi\rangle= a_i |\xi \rangle ~, ~~-{\partial \over
\partial \xi_i}|\xi\rangle=
a^\dagger_i |\xi\rangle, \label{Dala} \\
& \langle \xi | \xi^*_i =  \langle \xi | a^\dagger_i  ~, ~~
\langle \xi | {\partial \over \partial \xi^*_i}=  \langle \xi |
a_i , \label{Dalb}
\end{align}
\end{subequations}
it is easy to derive the equation of motion for the density
operator $\rho(t)$:
\begin{align}
\dot{\rho}(t)=\sum_{ij}\Big\{[&\dot{u_0}(t)u_0(t)^{-1}]_{ij}
a^{\dag}_{i}a_{j}\rho(t) \nonumber \\&
+[\dot{u_0}(t)u_0(t)^{-1}]^{*}_{ij}\rho(t) a^{\dag}_{j}a_{i}
\Big\} . \label{sle}
\end{align}
One can see from Eq.~(\ref{fsx}) that
$\dot{u_0}(t)u_0(t)^{-1}=-i\begin{pmatrix}E_{1}&T_c\\
T_c&E_{2}\end{pmatrix}$.  This simply reduces Eq.~(\ref{sle}) to
the familiar Liouvillian equation,
\begin{align}
\dot{\rho}(t)=-i[H,\rho(t)], \end{align} as we expected. Having
ensured the path integral technique can reproduce the dynamical
equation for an isolated double quantum dot, we will apply it to
the double quantum dot coupling to electron reservoirs in the next
section.

\section{The Master Equation for a Double Quantum Dot Gated by Electrodes}

A double quantum dot between two electron reservoirs, the source
and drain electrodes controlled via a bias voltage (see
Fig.~\ref{fig1}), has a total Hamiltonian as
\begin{eqnarray}
\label{hamil} H &=& H_{\rm dot}+H_{\rm rev}+H_I,
\end{eqnarray}
where
\begin{subequations}
\label{HM-0}
\begin{align}
H_{\rm dot}= E_1 a^\dagger_1 a_1 & + E_2 a^\dagger_2 a_2 +T_c(
a_{2}^{\dagger}a_{1}+a_{1}^{\dagger}a_{2}) \label{HM-0a}
\end{align} is the Hamiltonian of the double quantum dots,
\begin{align}
H_{\rm rev}=\sum_{k} (\varepsilon_{L k}a^{\dagger}_{L k}a_{L k} +
\varepsilon_{R k}a^{\dagger}_{R k}a_{R k} )  \label{HM-0b}
\end{align} is for the source and drain electrodes (reservoirs)
and
\begin{align}
H_I= \sum_{k}(t_{1Lk}a^{\dag}_{1}a_{Lk} +t_{2Rk} a^{\dag}_{R
k}a_{2} + {\rm H.c}) \label{HM-0c}
\end{align}
\end{subequations}
is the coupling (interaction) between the double dot and the
reservoirs that depicts the electron tunneling between them, where
subscript $k$ labels an electron state in the reservoirs, $L$ and
$R$ denote the reservoirs of the source (left) and drain (right)
electrodes, respectively, and $t_{1Lk} (t_{2Rk})$ is the electron
tunneling amplitude between the source (drain) and the left
(right) dot. Since we only concern in this work the dynamics of
charge qubit, we omitted spin degrees of freedom for electrons.

It should be pointed out that in Eq.~(\ref{HM-0a}) we did not
include explicitly the inter-dot Coulomb repulsion. This is
because a typical inter-dot Coulomb energy is of the order of
hundreds $\mu$eV, which is much larger than the energy level
splitting $\varepsilon$ and the inter-dot tunnel coupling $\Delta$
(both are of the order of tens $\mu$eV or less) for charge qubit
manipulation \cite{hayashi}. As a result, the inter-dot Coulomb
interaction simply leads one to exclude the states corresponding
to a simultaneous occupation of the two dots \cite{tb,Gurvitz}.
Thus, we will derive an exact master equation for the reduced
density matrix of the double dot without considering the inter-dot
Coulomb repulsion at beginning. The master equation of the double
dot in the strong inter-dot Coulomb repulsion regime is then
obtained from the exact master equation by explicitly excluding
the states of doubly occupied two dots in terms of Bloch-type rate
equations, as we will see later.

To derive non-puterbatively the master equation of the reduced
density matrix for the double dot system, we adopt the treatment
of the Feynman-Vernon influence functional theory \cite{fv63}.
This approach within the framework of path integral traces over
the degrees of freedom of the environment (here the electrodes)
into a functional of the dynamical variables of the system (the
double quantum dot). This functional is called the influence
functional by Feynman and Vernon, which contains all the dynamical
effects from the back-action of the reservoirs to the system due
to the coupling between them. Following the fermionic path
integral technique presented in the last section, we depict the
route to an exact master equation aided with the results derived
in details in the appendices.

\subsection{Influence functional}
Explicitly, the total density matrix of the double dot plus the
reservoirs obeys the
quantum Liouvillian equation $i \partial \rho _{\mathrm{tot}%
}(t)/\partial t=[H,\rho _{\mathrm{tot}}(t)]$, which yields the
formal solution:
\begin{equation}
\rho_{\mathrm{tot}}( t) =e^{-iH(t-t_0)}\rho_{\mathrm{tot}}(t_0)
e^{iH(t-t_0)}.
\end{equation}
As we are interested in dynamics of the electrons in the double
dot, we shall concentrate on the reduced density matrix for the
double dot by tracing out the environmental variables:
$\rho(t)={\rm tr}_E \rho_{\rm{tot}}(t)$. Assuming initially the
dots and the reservoirs are uncorrelated \cite{legt,hu}, $\rho
_{\mathrm{tot}}(t_0)=\rho (t_0)\otimes \rho _{E}(t_0)$, then the
reduced density matrix describing the full dynamics of the
electrons in the double dot becomes
\begin{align}
\langle\xi_f|\rho(t)|\eta_f\rangle=\int d\mu(\xi_0)& d\mu(\eta_0)
\langle\xi_0|\rho(t_0)|\eta_0\rangle \nonumber \\ & \times
J(\xi_f^*,\eta_f,t|\xi_0,\eta_0^*,t_0), \label{rdm}
\end{align} where the propagating function is
defined as
\begin{align}
J(\xi_f^*,&\eta_f,t|\xi_0,\eta_0^*,t_0)=\nonumber\\
&\int\mathcal{D}[\xi^*\xi;\eta^*\eta]e^{i
(S_{c}[\xi^*,~\xi]-S^*_{c}[\eta^*,~\eta])}\mathcal{F}
[\xi^*\xi;\eta^*\eta] \label{ppg}
\end{align}
with $S_{c}[\xi^*,\xi]$ being the action of the double dot in the
fermion coherent state representation given by Eq.~(\ref{action}),
and $\mathcal{F}[\xi^*\xi;\eta^*\eta]$ is the influence functional
which takes fully into account the back-action effects of the
reservoirs to the double dot and modifies the original action of
the system into an effective one,
$e^{{i\over\hbar}(S_{c}[\xi^*,\xi]-
S^*_{c}[\eta^*,\eta])}\mathcal{F}[\xi^*\xi;\eta^*\eta]
=e^{{i\over\hbar}S_{eff}[\xi^*\xi;\eta^*\eta] }$. The path
integral $\mathcal{D}[\xi^*\xi;\eta^*\eta]$ integrates over all
pathes $\xi^*(\tau),\xi(\tau),\eta^*(\tau)$, and $\eta(\tau)$ in
the Grassmann space bounded by $\xi^*(t)=\xi^*_{f}$,
$\xi(t_0)=\xi_0$, $\eta^*(t_0)=\eta^*_0$, and $\eta(t)=\eta_{f}$.

Let the reservoirs be initially in thermal equilibrium states at
temperature $\beta^{-1}$, the influence functional can then be
solved exactly with the result (see the derivation in the Appendix
A):
\begin{widetext}
\begin{align}
\mathcal{F}_{}[\xi^*\xi;\eta^*\eta]=& \exp \sum_{i=1,2}
\Big\{-\int_{t_0}^{t}d\tau \int_{t_0}^{\tau}d\tau'\Big(F_{il}
(\tau-\tau')\xi^*_{i}(\tau)\xi_{i}(\tau')
+F^{*}_{il}(\tau-\tau')\eta^*_{i}(\tau')\eta_{i}(\tau)\Big)
\nonumber\\ & +\int_{t_0}^{t}d\tau\int_{t_0}^{t}d\tau'\Big(
F_{il}(\tau-\tau') \eta^*_{i}(\tau) \xi_{i}(\tau')
-F_{il}^\beta(\tau-\tau') \big(\eta^*_{i}(\tau)
+\xi^*_{i}(\tau)\big)\big(\eta_{i}(\tau') +\xi_{i}(\tau')\big)
\Big)\Big\}. \label{inf}
\end{align}
\end{widetext}
The two time correlation functions in the influence functional,
\begin{subequations}
\label{Kernel}
\begin{align}
&F_{il}(\tau-\tau')=\sum_{k}|t_{il k}|^2e^{-i\varepsilon_{l
k}(\tau-\tau')},
\label{Kernela} \\
&F_{il}^{\beta}(\tau-\tau')=\sum_{k} f(\varepsilon_{l k})|t_{il
k}|^2 e^{-i\varepsilon_{l k } (\tau-\tau')} ,\label{Kernelb}
\end{align}
\end{subequations}
are called the dissipation-fluctuation kernels, where
$f(\varepsilon_{l k})={1\over e^{\beta(\varepsilon_{l
k}-\mu_{l})}+1}$ with $l=L,R$ for $i=1,2$, respectively, are the
fermi distribution functions of the electron reservoirs, and
$\mu_{L,R}$ are the corresponding chemical potentials. These
nonlocal time-dependent functions contains the full dynamics
effect from the reservoirs to the double dot. We should also point
out that in the coherent state representation, the influence
functional for a many-fermion environment has a form similar to
that of a many-boson environment \cite{an07} except for some sign
difference due to the antisymmetric properties of fermion degrees
of freedom.

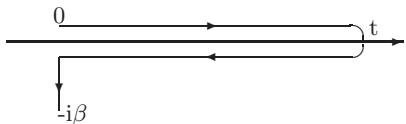
\begin{figure}[ht]
\begin{picture}(200,40)(10,20)
\put (60,50){\makebox(0,0){0}} \put (179,45){\makebox(0,0){t}}
\put (40,40){\vector(1,0){150}} \put (60,46){\vector(1,0){60}}
\put (60,46){\line(1,0){110}} \put(170,41) {\oval(10,10.2)[tr]}
\put(170,39) {\oval(10,10)[br]} \put (60,34){\line(1,0){110}} \put
(170,34){\vector(-1,0){55}} \put (60,34){\line(0,-1){20}} \put
(60,34){\vector(0,-1){14}} \put (65,12){\makebox(0,0){-i$\beta$}}
\end{picture}
\caption{The closed-time path for the trace of the density matrix
of the reservoirs.} \label{fig2}
\end{figure}

The physical meaning of the above influence functional is very
clear. The four terms contained in the exponential function of
Eq.~(\ref{inf}) correspond to four different physical processes in
a time-closed path formulism, see Fig.~\ref{fig2}.
The first term gives the contribution of the back-action effect of
the reservoirs to the double dot in terms of the time correlation
function $F_{il}(\tau-\tau')$ of Eq.~(\ref{Kernela}) in the
forward process from the time $t_0$ to the time $t$. For the
double time integrals it contains, the first one starts from $t_0$
to $\tau$ and the second one from $t_0$ to $t$, sums over all the
time sequences of the these propagations. Resummation of these
propagating processes up to all orders of the system-reservoir
coupling results in exactly the exponential function appearing in
Eq.~(\ref{inf}). Similar to the first term, the second term are
just the back-action effect of the reservoirs to the system in the
backward process from the time $t$ back to the time $t_0$ which is
just the complex conjugate of the first term. In terms of the
Schwinger-Keldysh Green-function approach, the time correlation
function in these two propagating processes are just the
time-ordered and anti-time ordered Green functions. The histories
of the forward pathes $\xi^*_{i}(\tau),\xi_{i}(\tau)$ and the
backward pathes $\eta^*_{i}(\tau),\eta_{i}(\tau)$ are mixed up
through the time correlation functions as shown in the third and
the fourth terms in Eq.~(\ref{inf}). The third term in
Eq.~(\ref{inf}) represents the mix of the forward and backward
pathes at the time $t$. The last term is a mix of the forward and
backward pathes at the time $t_0$ where the initial equilibrium
properties of the reservoirs, i.e. the fermi statistics and the
temperature of the reservoirs, naturally enter into the time
correlation function $F_{il}^{\beta}(\tau-\tau')$, as shown by
Eq.~({\ref{Kernelb}).

As we see all the influences of the reservoirs on the double dot
are embedded in the two time-correlation functions, i.e., the two
dissipation-fluctuation kernels (\ref{Kernel}) in the influence
functional. These two dissipation-fluctuation kernels are related
each other through the dissipation-fluctuation theorem. This will
become clear by introducing a spectral density defined as
\begin{align} \label{spectral}
J_{il}(\omega)=\sum_{k}|t_{il k} |^2 \delta(\omega-\varepsilon_{l
k}).
\end{align}
Obviously, the spectral density contains all the information about
the reservoirs' density of states involved in the electron
tunneling between the dots and reservoirs. Then the
time-correlation functions (\ref{Kernel}) can be expressed as
\begin{subequations}
\label{Kernels}
\begin{align}
&F_{il}(\tau-\tau')=\int d\omega
J_{il}(\omega)e^{-i\omega(\tau-\tau')},
\label{Kernelsa} \\
&F_{il}^{\beta}(\tau-\tau')=\int d\omega
{J_{il}(\omega)e^{-i\omega(\tau-\tau')} \over e^{
\beta(\omega-\mu_l)} +1},\label{Kernelsb}
\end{align}
\end{subequations}
This equation manifests the dissipation-fluctuation theorem in an
open quantum system. It tells that all the back-action effects of
the reservoirs to the double dot are crucially determined by the
spectral density $J_{il}(\omega)$.

\subsection{Exact master equation}
Now we can derive the master equation for the reduced density
matrix.  As we see the effective action after integrating out the
environmental degrees of freedom, i.e. combining Eqs.~(\ref{ppg})
and (\ref{inf}) together, still has a quadratic form in terms of
the dynamical variables of the fermion coherent states. Thus the
path integral (\ref{ppg}) can be solved exactly by utilizing the
stationary path method and gaussian integrals
\cite{fey65,Faddeev}. The resulting propagating function is simply
given by
\begin{widetext}
\begin{align}
J(\xi_f^*,\eta_f,t|\xi_0,\eta_0^*,t_0)=A(t)\exp \sum_{i=1,2}\Big[
{\xi^*_{i f}\xi_{i}(t)+\xi^{*}_{i}(t_0)\xi_{i 0}\over2}
+{\eta^*_{i 0}\eta_{i}(t_0) +\eta^{*}_{i}(t)\eta_{i f}\over2}\Big]
\label{ppgs}.
\end{align}
where $A(t)$ is the contribution arisen from the fluctuations
around the stationary pathes which will be given later. The
stationary pathes $\xi_{i}(t)$ and $\eta_{i}(t_0)$ are determined
by the equations of motion
\begin{subequations}
\label{EM-1}
\begin{align}
\dot{\xi}_i(\tau)+i[E_i \xi_i(\tau) + T_c\xi_{i'}(\tau)]
+\int_{t_0}^{\tau}d\tau'F_{il} (\tau-\tau')\xi_i
(\tau')&-\int_{t_0}^{t}d\tau' F_{il}^{\beta}(\tau-\tau')
[\xi_i(\tau')+\eta_i(\tau')] =0 \label{EM-1a}
\\
\dot{\eta}_i (\tau)+i [E_i \eta_i(\tau)+ T_c \eta_{i'}(\tau)]
+\int_{t_0}^{\tau}d\tau' F_{il}(\tau-\tau')
\eta_i(\tau')&-\int_{t_0}^{t}d\tau' F_{il}(\tau-\tau')
[\eta_i(\tau')+\xi_i(\tau')] \nonumber \\
& + \int_{t_0}^{t}d\tau' F_{il}^{\beta}(\tau-\tau')
[\eta_i(\tau')+\xi_i(\tau')] =0 \label{EM-1b}
\end{align}
\end{subequations}
\end{widetext}
subjected to the boundary condition $\xi_i (t_0) = \xi_{i 0}$ and
$\eta_{i}(t)=\eta_{i f}$, while $\xi^*_{i}(t_0)$ and
$\eta^{*}_{i}(t)$ in Eq.~(\ref{ppgs}) can be obtained from the
complex conjugate equations of (\ref{EM-1b}) and (\ref{EM-1a}),
respectively, under the boundary condition $\xi^*_i (t) = \xi^*_{i
f}$ and $\eta^*_{i}(t_0)=\eta^*_{i 0}$. The local terms
$\dot{\xi}_i+i(E_i \xi_i + T_c\xi_{i'})$ and $\dot{\eta}_i +i(E_i
\eta_i+T_c\eta_{i'})$ in Eq.~(\ref{EM-1}) are intrinsic and well
describe the coherent dynamics of the electron states in an
isolated double quantum dot, as we have discussed in the last
section. The nonlocal terms involving two different time
correlation functions, $F_{il}(\tau-\tau')$ and
$F_{il}^{\beta}(\tau-\tau')$, stem from the coupling to the
reservoirs. These two time correlation functions play quite
distinct roles in the equations of motion. The interaction between
the double dot and the electron reservoirs is mediated through
electron tunnelings between them. The correlation
$F_{il}(\tau-\tau')$ describes the back-action of the reservoirs
to the double dot due to the interaction between them. However,
the correlation function $F_{il}^{\beta}(\tau-\tau')$ also harbors
the Fermi-Dirac statistic effect of the electron reservoirs which
exists even in the zero-temperature limit. The latter situation is
quite different from a bosonic environment where
$F_{il}^{\beta}(\tau-\tau')$ vanishes at zero temperature
\cite{an07}.

The solutions of $\xi_{i}(t)$, $\eta_{i}(t_0)$ and
$\eta^{*}_{i}(t)$, $\xi^{*}_{i}(t_0)$ determined by
Eq.~(\ref{EM-1}) and its complex conjugate can be factorized from
the corresponding boundary conditions. It is not too difficult to
find that\begin{subequations} \label{SEM-1}
\begin{align}
& \xi(t)=(I-v(t))^{-1}\big[u(t)\xi_{0}+v(t)\eta_{f}\big],~ \\
& \eta(t_0)=u^{\dag}(t)[I+(I-v(t))^{-1}v(t)]\eta_{f} \nonumber
\\&~~~~~~~~~~~~~~
-\big[I-u^\dagger(t)(I-v(t))^{-1}u(t)\big]\xi_{0}.
\end{align}
\end{subequations}
Here we have again used the matrix notation: $\xi^T=(\xi_{1}
~\xi_{2})$ and a same form for $\eta$. The new dynamical variables
expressed as two time-dependent $2 \times 2$ matrices, $u(\tau)$
and $v(\tau)$, satisfy the dissipation-fluctuation
integrodifferential equations:
\begin{subequations}
\label{EM-3}
\begin{align}
\dot{u}(\tau)+iM u(\tau) +\int^{\tau}_{t_0}& d\tau'
 G(\tau-\tau')u(\tau')=0 ,  \label{EM-3a} \\
\dot{v}(\tau)+iM v(\tau) +\int_{t_0}^{\tau}& d\tau' G(\tau-\tau')
v^{}(\tau') \nonumber \\ & = \int_{t_0}^{t} d\tau'
G^\beta(\tau-\tau') \bar{u}(\tau'), \label{EM-3b}
\end{align}
\end{subequations}
with the boundary conditions $u_{ij}(t_0)=\delta_{ij}$ and
$v_{ij}(t_0)=0$, while $\bar{u}(\tau)$ in Eq.~(\ref{EM-3b}) obeys
the backward equation of motion to $u(\tau)$, namely,
$\bar{u}(\tau)=u^\dagger(t+t_0-\tau)$ for $t_0\le \tau \le t$. In
Eq.~(\ref{EM-3}), we have also defined the $2\times 2$ matrices
$M_{ij}= E_{i} (i=j)$ and $T_c (i\neq j)$, $G_{ij}(\tau-\tau')=
F_{il}(\tau-\tau')\delta_{ij}$, and $G^\beta_{ij}(\tau-\tau')=
F_{il}^{\beta} (\tau-\tau') \delta_{ij}$. As we see $u(\tau)$ is
determined purely by $F_{il}(\tau-\tau')$ while $v(\tau)$ depends
on both correlation functions. The full complexity of the
non-Markovian dynamics of charge coherence in the double dot
induced from its coupling to the reservoirs is thus manifested
through these equations of motion.

As $\xi^*_{i}(t_0)$ and $\eta^{*}_{i}(t)$ satisfying respectively
the complex conjugate equations of $\eta_{i}(\tau)$ and
$\xi_{i}(\tau)$ with the boundary condition $\xi^*_i (t) =
\xi^*_{i f}$ and $\eta^*_{i}(t_0)=\eta^*_{i 0}$, it is easy to
find the similar solution from Eq.~(\ref{SEM-1}) for
$\xi^*_{i}(t_0)$ and $\eta^{*}_{i}(t)$. Substituting these results
into Eq.~(\ref{ppgs}) and note the fact that $v(\tau)$ is a
hermitian matrix at $\tau=t$, we obtain explicitly the exact
propagating function of the double dot system:
\begin{align}
J(\xi^{*}_{f}, \eta_{f},t|& \xi_{0},\eta^{*}_{0},t_0)= A(t) \exp
\Big\{\xi^\dagger_f J_1 (t)\xi_0  \nonumber
\\ & + \xi^\dagger_f J_2(t)\eta_f + \eta^\dagger_0 J_3(t)\xi_0
+ \eta^\dagger_0 J^\dagger_1(t) \eta_f \Big\} , \label{ppg1}
\end{align}
where
\begin{subequations}
\label{coef0}
\begin{align}
&J_1(t)=w(t)u(t), ~~
J_2(t)=w(t)-I, \\
&J_3(t)=u^{\dag}(t)w(t)u(t)-I,~~ A(t)=1/{\rm det}(w(t)).
\end{align}
\end{subequations}
with $w(t)=(I-v(t))^{-1}$. All these time-dependent coefficients
can be fully determined by solving Eq.~(\ref{EM-3}).

Once the exact propagating function is obtained, the dynamics of the
reduced density matrix, Eq.~(\ref{rdm}), which takes fully into
account the back-action of the electron reservoirs, can be
completely solved for any given initial electron state of the double
dot. The explicit solution of the reduced density matrix relies
solely on the solution to the equations of motion (\ref{EM-3})
which, in general, has to be solved numerically. To check the
consistency, we may let the double dot be decoupled from the
electron reservoirs, namely, set $t_{il k}=0$, then the
dissipation-fluctuation kernels vanish. As a result,
Eq.~(\ref{EM-3a}) is reduced to Eq.~(\ref{fsx}) and $u(t)$ is
reduced to $u_0(t)$ whose solution is given after Eq.~(\ref{fsx}) in
the last section, while the solution of Eq.~(\ref{EM-3b}) gives
$v(t)=0$. Consequently the propagating function Eq.~(\ref{ppg1}) is
reduced to Eq.~(\ref{ppg0s}) which recovers the exact solution of
the isolated double dot system shown in the last section.

Having the explicit form of the propagating function (\ref{ppg1})
in hands, it is straightforward to derive the master equation for
the reduced density matrix directly from Eq.~(\ref{rdm}). Here we
shall deduce an operator form of the master equation such that all
the time-dependent coefficients in the master equation are
explicitly independent from the initial state of the double dot as
well as from any specific representation. Taking the time
derivative to Eq.~(\ref{ppg1}), eliminating the initial state
dependence and using the D-algebra of the fermion creation and
annihilation operators in the fermion coherent state
representation, Eq.(\ref{Dal}), the exact master equation of the
double dot with a general spectral density at arbitrary
temperature and bias is given by:
\begin{align}
\dot{\rho}(t)=&\sum_{ij}\Big\{\Omega_{ij}(t)[a^\dag_ia_{j},
\rho(t)] \nonumber \\
& + \Gamma_{ij}(t)(2a_{j}\rho(t) a^{\dag}_{i}-
a^{\dag}_{i}a_{j}\rho(t)-\rho(t) a^{\dag}_{i}a_{j}) \nonumber\\&
+\Gamma^\beta_{ij}(t)(a_{j}\rho(t) a^{\dag}_{i} -
a^{\dag}_{i}\rho(t) a_{j}- a^{\dag}_{i}a_{j}\rho(t)-\rho(t)
a^{\dag}_{i}a_{j})\Big\} \nonumber \\ &+\Gamma^0(t)\rho(t) ,
\label{emaster}
\end{align}
where all time-dependent coefficients in Eqs.~(\ref{emaster}) are
determined by $u(t)$ and $v(t)$ through the following relations:
\begin{subequations}
\label{td-coe}
\begin{align}
&\Omega_{ij}(t)= {1\over 2}[\dot{u}u^{-1}-
(u^\dag)^{-1}\dot{u}^\dag]_{ij}, \\
&\Gamma_{ij}(t)= -{1\over 2}[\dot{u}u^{-1}+
(u^\dag)^{-1}\dot{u}^\dag]_{ij}, \\
&\Gamma^\beta_{ij}(t)=[\dot{u}u^{-1}v
+v(u^\dag)^{-1}\dot{u}^\dag-\dot{v}]_{ij},
\end{align}
\end{subequations}
and $\Gamma^0(t)={\rm tr}\Gamma^\beta$.
 The first term in the master equation is indeed the
generalized Liouvillian term which can be explicitly written as
\begin{align}
\sum_{ij}\Omega_{ij}(t)[a^\dag_ia_{j}, \rho ]=-i[H'(t),\rho]
\end{align} where
\begin{align} \label{rh}
H'(t)&=E'_{1}(t)a^{\dag}_{1}a_{1}+E'_{2}(t) a^{\dag}_{2}a_{2}+
{T'_c(t)}a^{\dag}_{1}a_{2}+{T'^*_c(t)}a^{\dag}_{2}a_{1}
\end{align} is an effective Hamiltonian of the double quantum dot
with the shifted (renormalized) time-dependent energy levels and
the shifted inter-dot transition amplitude,
\begin{subequations}
\label{ec-ren}
\begin{align}
&E'_{i}(t)= i \Omega_{ii}~~(i=1,2),~~~ T'_c(t)= i\Omega_{12} .
\end{align}
Using the equation of motion (\ref{EM-3a}), we further obtain
\begin{align}
&E'_{i}(t)= E_i -
 \mbox{Im}[W_{ii}(t)],~~i=1,2 , \\
&T'_c(t)=T_c +{i\over 2}[W_{12}(t)-W^*_{21}(t)],
\end{align}
where
\begin{align}
W(t)=\int_{t_0}^{t}d\tau
\begin{pmatrix}F_{1L}(t-\tau)&0\\0&F_{2R}(t-\tau)\end{pmatrix}u^{}(\tau)
u(t)^{-1} ,
\end{align}
\end{subequations}
It shows that the shifted energy levels, $E'_{1,2}(t)$, and the
inter-dot transition amplitude, $T'_c(t)$, are entirely contributed
by $u(t)$ that involves only the time-correlation function
$F_{il}(t-\tau)$. The rest part in Eq.~(\ref{emaster}) describes the
dissipation and noise processes of electron charges with
non-Markovian behaviors having been embedded in these time-dependent
$2\times 2$ matrix coefficients, $\Gamma(t)$ and $\Gamma^\beta(t)$.
We call $\Gamma(t)$ and $\Gamma^\beta(t)$ the
dissipation-fluctuation matrix coefficients or simply the
dissipation-fluctuation coefficients hereafter. Note that
$\Gamma(t)$ is solely determined by $u(t)$, and $\Gamma^\beta(t)$ is
given by both $u(t)$ and $v(t)$. Thus all the time-dependent
coefficients in the master equation are non-perturbatively
determined by Eq.~(\ref{EM-3}) and fully account the back-action
effects of the reservoirs to the double dot.

\subsection{Bloch-type Rate equations for zero and strong
inter-dot Coulomb repulsion cases and the corresponding Markovian
limits}

\indent {\it (a) No inter-dot Coulomb repulsion}:  To closely
examine the decoherence of electron charge dynamics in the double
dot, it is more convenient to rewrite the master equation
(\ref{emaster}) as a set of Bloch-type rate equations in terms of
the localized charge states in the double dot. Without considering
the inter-dot Coulomb repulsion, the rate equations can be
obtained directly from the master equation (\ref{emaster}) in the
charge configuration space containing the states of empty double
dot, the first dot occupied, the second dot occupied and both dots
occupied. We label these four states by $|j\rangle, j=0,1,2,3$,
respectively. Then the master equation in the above basis becomes
\begin{subequations}
\label{ni-re}
\begin{align}
\dot{\rho}_{00}&=\tilde{\Gamma}_{11}\rho_{11} +
\tilde{\Gamma}_{21}\rho_{12} + \tilde{\Gamma}_{12}\rho_{21} +
\tilde{\Gamma}_{22}\rho_{22}+\Gamma^0 \rho_{00}, \\
\dot{\rho}_{11}&=(\Gamma^0-2\bar{\Gamma}_{11})\rho_{11}
+\bar{\Xi}^*_- \rho_{12}+\bar{\Xi}_- \rho_{21} \nonumber \\ &
~~~~~~~~~~~~~~~~~~~~~~~~~~
-\Gamma^\beta_{11}\rho_{00}+\tilde{\Gamma}_{22}\rho_{33},
 \\ \dot{\rho}_{22}&=(\Gamma^0-2\bar{\Gamma}_{22})\rho_{22}
+\bar{\Xi}^*_+ \rho_{12} +\bar{\Xi}_+\rho_{21} \nonumber
\\ & ~~~~~~~~~~~~~~~~~~~~~~~~~~
-\Gamma^\beta_{22}\rho_{00}+\tilde{\Gamma}_{11}\rho_{33}, \\
\dot{\rho}_{12}&=[-i\varepsilon' + \Gamma^0 -{\rm
tr}\bar{\Gamma}]\rho_{12} +  \bar{\Xi}_+ \rho_{11} +
\bar{\Xi}_-\rho_{22} \nonumber
\\ & ~~~~~~~~~~~~~~~~~~~~~~~~~~
-\Gamma^\beta_{12}\rho_{00}+\tilde{\Gamma}_{12}\rho_{33}, \\
\dot{\rho}_{33}&=\Gamma^\beta_{12}\rho_{21}+\Gamma^\beta_{21}\rho_{12}
-\Gamma^\beta_{11}\rho_{22}-\Gamma^\beta_{22}\rho_{11} \nonumber
\\ &~~~~~~~~~~~~~~~~~~~~~~~~~~ +(\Gamma^0-2{\rm
tr}\bar{\Gamma})\rho_{33}.
\end{align}
\end{subequations}
Here the density matrix elements are defined by $\rho_{ij}=\langle
i|\rho|j\rangle$. We have also defined all the time-dependent
coefficients in the rate equations as
$\tilde{\Gamma}(t)=2\Gamma(t)+\Gamma^\beta(t)$,
$\bar{\Gamma}(t)=\Gamma(t)+\Gamma^\beta(t)$, $\bar{\Xi}_{\pm}=\pm
i T'_c(t) - \bar{\Gamma}_{12}$ and $\varepsilon'(t)=
E'_1(t)-E'_2(t)$ where $E'_i(t),T'_c(t), \Gamma(t),
\Gamma^\beta(t)$ and $\Gamma^0(t)$ are the time-dependent
transport coefficients contained in the master equation and are
explicitly given by Eqs.~(\ref{td-coe}) and (\ref{ec-ren}). These
time-dependent coefficients in the rate equations thus fully
characterize the non-Markovian dynamics of electrons in the double
dot. The first and the last rate equations account electron charge
leakage effects in this device, while other three rate equations
depict charge qubit decoherence dynamics under the influence of
the reservoirs.

For a constant spectral density that has been widely used in the
literature, the time-correlation function becomes
\begin{align}
F_{il}(\tau-\tau') \rightarrow
{\Gamma_{l}\over2\pi}\int_{-\infty}^{\infty}d\omega
e^{-i\omega(\tau-\tau')}=\Gamma_{l}\delta(\tau-\tau'),
\end{align}
where $\Gamma_{l}=2\pi \rho_{l}|t_{ilk}|^2$, ($l=L,R$ for $i=1,2$),
and $\rho_{L,R}$ are the densities of states for the left and right
electron reservoirs. Then in the Markovian limit ($t-t_0$, $\tau-t_0
\rightarrow \infty$) \cite{Carmichael93}, the integral kernels in
Eq.~(\ref{EM-3}) reduce to
\begin{subequations}
\label{dfc}
\begin{align} &\int_{t_0}^{\tau}d\tau'
F_{il}(\tau-\tau') u_{ij}(\tau')
\rightarrow {1\over2}\Gamma_{l}u_{ij}(\tau), \label{dfc-a}\\
&\int_{t_{0}}^{t} d\tau'
F^{\beta}_{il}(\tau-\tau')\bar{u}_{ij}(\tau') \rightarrow
f_{l}\Gamma_{l}\bar{u}_{ij}(\tau) . \label{dfc-b}
\end{align}
\end{subequations}
The factor ${1\over2}$ in Eq.(\ref{dfc-a}) comes from the boundary
of the time integration sitting upon $\tau$, and $f_{L,R}$ in
Eq.~(\ref{dfc-b}) are the fermi distribution functions of the
reservoirs. In this Markovian limit, the equation of motion
(\ref{EM-3}) can be solved analytically and the solution is
\begin{subequations}
\label{sol-dfe}
\begin{align}
&u(\tau)=\exp \Big\{ -\begin{pmatrix}iE_{1}+{\Gamma_{L}\over2}&iT_c \\
iT_c &iE_{2} +{\Gamma_{R}\over2}
\end{pmatrix}(\tau-t_0)\Big\}, \\
&v(\tau)=\int_{t_{0}}^{\tau}d\tau' u(\tau+t_0-\tau')
\begin{pmatrix}f_{L}\Gamma_{L}&0\\
0&f_{R}{\Gamma_{R}}\end{pmatrix}\bar{u}(\tau'+t_0).
\end{align}
\end{subequations}

With the above explicit solution for $u(t)$ and $v(t)$, we can
obtain analytically all the coefficients in the master equation as
well. Note that $\bar{u}(\tau)=u^\dagger(t+t_0-\tau)$ depending
explicitly on the time $t$  implies that
$\partial_{\tau}v(\tau)|_{\tau=t} \ne
\partial_{t}v(t) = \dot{v}(t)$, it is easy to find
\begin{subequations}
\label{tid-coe0}
\begin{align}
&\dot{u}u^{-1}= -\begin{pmatrix}iE_{1}+{\Gamma_{L}\over2}&iT_c \\
iT_c &iE_{2} +{\Gamma_{R}\over2}
\end{pmatrix}, \\
&\dot{u}u^{-1}v +v(u^\dag)^{-1}\dot{u}^\dag-\dot{v}=-
\begin{pmatrix}f_{L}\Gamma_{L}&0\\
0&f_{R}{\Gamma_{R}}\end{pmatrix} .
\end{align}
\end{subequations}
As a result, for a constant spectral density in the Markovian
limit, all the time-dependent coefficients in the master equation
become constant:
\begin{subequations}
\label{tid-coe}
\begin{align}
&E'_{1,2}(t) \rightarrow E_{1,2}, ~~T'_c(t)
\rightarrow T_c , \\
&\Gamma_{ij}(t) \rightarrow {1\over 2}\Gamma_l\delta_{ij},
~~\Gamma^\beta_{ij}(t) \rightarrow -f_l\Gamma_l\delta_{ij} .
\end{align}
\end{subequations}
In other words, in the Markovian limit with a constant spectral
density, there is no renormalization effect to the energy level
shift and the inter-dot transition amplitude. The
dissipation-fluctuation effects are simply reduced to the
time-independent tunneling rates between the dots and the
reservoirs. In fact, the differences of time-dependent and
time-independent coefficients manifest the non-Markovian dynamics
of electron charges in the double dot. We shall present
quantitatively such differences in the next section.

For a constant spectral density in the Markovian limit, the rate
equations are simply reduced to
\begin{subequations}
\label{rate-2}
\begin{align}
\dot{\rho}_{00}&=\bar{f}_L\Gamma_L\rho_{11}+
\bar{f}_R\Gamma_R\rho_{22} -(f_L\Gamma_L+f_R\Gamma_R)\rho_{00} , \\
\dot{\rho}_{11} &=-(\bar{f}_L\Gamma_L+f_R\Gamma_R)\rho_{11}
+iT_c(\rho_{12}-\rho_{21})\nonumber \\
&~~~~~~~~~~~~~~+f_L\Gamma_L\rho_{00}+ \bar{f}_R\Gamma_R\rho_{33}, \\
\dot{\rho}_{22}& =-(\bar{f}_R\Gamma_R+f_{L}\Gamma_L)\rho_{22}
-iT_c(\rho_{12}-\rho_{21}) \nonumber \\
&~~~~~~~~~~~~~~+f_R\Gamma_R\rho_{00}+ \bar{f}_L\Gamma_L\rho_{33} ,
\\ \dot{\rho}_{12}&=(-i\varepsilon-{\Gamma_L+\Gamma_R\over 2})
\rho_{12}+iT_c(\rho_{11}-\rho_{22}),
\\ \dot{\rho}_{33}&= -(\bar{f}_L\Gamma_L+\bar{f}_R\Gamma_R]\rho_{33}
+f_L\Gamma_L\rho_{22} +f_R\Gamma_R\rho_{11},
\end{align}
\end{subequations}
where $\bar{f}_{L,R}=(1-f_{L,R})$ and $\varepsilon=E_1-E_2$. Under
the large bias limit, $f_L=1,~f_R=0$, the above rate equations
reproduce the rate equations obtained by Gurvitz and Prager
\cite{Gurvitz} for the double dot without considering the
inter-dot Coulomb repulsion:
\begin{subequations}
\label{gurvitzr}
\begin{align}
&\dot{\rho}_{00}=-\Gamma_L\rho_{00}+\Gamma_R\rho_{22},
\\& \dot{\rho}_{11}=\Gamma_L\rho_{00}+\Gamma_R\rho_{33}
+iT_c(\rho_{12}-\rho_{21}),\\&
\dot{\rho}_{22}=-(\Gamma_L+\Gamma_R)\rho_{22}
-iT_c(\rho_{12}-\rho_{21}), \\& \dot{\rho}_{12}=(-i\varepsilon
-{\Gamma_L+\Gamma_R\over2})\rho_{12}
+iT_c(\rho_{11}-\rho_{22}),\\&
\dot{\rho}_{33}=-\Gamma_R\rho_{33}+\Gamma_L\rho_{22}.
\end{align}
\end{subequations}

\indent {\it (b) Strong inter-dot Coulomb repulsion}: On the other
hand, realistic experiments of the double dot are set up in the
strong Coulomb blockade regime where not only each of dots has
only one effective energy level but also there is no states of
simultaneous occupation of the two dots. In other words, the
configuration space of the localized charge states in the double
dot system with a strong inter-dot Coulomb repulsion only contains
the states of empty double dot, the first dot occupied and the
second dot occupied, denoted by $|j\rangle, j=0,1,2,$
respectively. The corresponding rate equations in this strong
Coulomb blockade regime for an arbitrary spectral density can also
be obtained by simply excluding the doubly occupied states in
Eq.~(\ref{ni-re}). Since the rate of the doubly occupied state
$|3\rangle$ in the case of ignoring inter-dot Coulomb repulsion
depends on the populations $\rho_{11}$ and $\rho_{22}$. This
probability flow from the states $|1\rangle$ and $|2\rangle$ to
$|3\rangle$ should be redirected back into the states $|1\rangle$
and $|2\rangle$ in the strong inter-dot Coulomb repulsion regime.
Meanwhile, to ensure the probability conservation without the
doubly occupied states, a correction to the dependence of the
coherence elements $\rho_{12}$ and $\rho_{21}$ in the rate
equations for $\rho_{11}$ and $\rho_{22}$ must also be taken into
account guided by the condition:
\begin{align}
\Gamma^\beta_{12}\rho_{21}+\Gamma^\beta_{21}\rho_{12}
-\Gamma^\beta_{11}\rho_{22}-\Gamma^\beta_{22}\rho_{11} =0.
\end{align}
This condition indeed forces the doubly occupied state to decouple
from other states in the double dot, as one can see from the rate
equation $\dot{\rho}_{33}$ in Eq.~(\ref{ni-re}). These
modifications can be done explicitly by taking the following shift
to the coefficients in the non-interacting rate equations
(\ref{ni-re}): $\bar{\Gamma}_{ii} \rightarrow \bar{\Gamma}_{ii} +
{1\over 2}\Gamma^\beta_{jj}$, $\bar{\Gamma}_{ij} \rightarrow
\bar{\Gamma}_{ij} - {1\over 2}\Gamma^\beta_{ij}$ with $i\ne j$. In
fact, the above coefficient shift also automatically cancels the
$\Gamma^0$ dependence of $\rho_{33}$ in Eq.~(\ref{ni-re}), which
is indeed a criterion for entirely excluding the double occupied
state from the reduced density matrix, as one can directly see
from the expression of the master equation (\ref{emaster}). Then
the rate equation of $\rho_{33}$ in Eq.~(\ref{ni-re}) becomes
$\dot{\rho}_{33}=(-2{\rm tr}\bar{\Gamma})\rho_{33}$, its solution
is $\rho_{33}(t)=0$ if initially $\rho_{33}(t_0)=0$. This implies
that no leakage into the double occupied state will occur. As a
result, the rate equations in the strong inter-dot Coulomb
repulsion regime are given by
\begin{subequations}
\label{i-re}
\begin{align}
\dot{\rho}_{00}&=\tilde{\Gamma}_{11}\rho_{11} +
\tilde{\Gamma}_{21}\rho_{12} + \tilde{\Gamma}_{12}\rho_{21} +
\tilde{\Gamma}_{22}\rho_{22}+\Gamma^0 \rho_{00}, \\
\dot{\rho}_{11}&=-\tilde{\Gamma}_{11}\rho_{11}
+\tilde{\Xi}^*_{-}\rho_{12} +\tilde{\Xi}_{-}\rho_{21}
-\Gamma^\beta_{11}\rho_{00},
 \\ \dot{\rho}_{22}&=-\tilde{\Gamma}_{22}\rho_{22}
+\tilde{\Xi}^*_{+}\rho_{12} +\tilde{\Xi}_{+}\rho_{21}
-\Gamma^\beta_{22}\rho_{00}, \\
\dot{\rho}_{12}&=[-i\varepsilon' -{1\over 2}{\rm
tr}\tilde{\Gamma}]\rho_{12} +\tilde{\Xi}_+\rho_{11}
+\tilde{\Xi}_-\rho_{22} -\Gamma^\beta_{12}\rho_{00},
\end{align}
\end{subequations}
where $\tilde{\Xi}_\pm(t)=\pm iT'_c(t)-{1\over2}
\tilde{\Gamma}_{12}(t)$. This set of the rate equations depicts the
full non-Markovian dynamics of the double dot in the strong
inter-dot Coulomb repulsion regime.

For the case of a constant spectral density in the Markovian
limit, $\tilde{\Gamma}_{ij}(t)=(2\Gamma(t)+\Gamma^\beta(t))_{ij}
\rightarrow \bar{f}_l\Gamma_l \delta_{ij}$, $\tilde{\Xi}_\pm (t)
\rightarrow \pm i T_c$ and $\varepsilon'(t) \rightarrow
\varepsilon$. The rate equations under such circumstances are
reduced to the rate equations for the double dot in the strong
inter-dot Coulomb repulsion regime, given in \cite{tb},
\begin{subequations}
\label{i-re1}
\begin{align}
&\dot{\rho}_{00}=-(f_L\Gamma_L+f_R\Gamma_R)\rho_{00}+
\bar{f}_L\Gamma_L\rho_{11} +\bar{f}_R\Gamma_R\rho_{22},
 \\& \dot{\rho}_{11}=f_L\Gamma_L\rho_{00}
+iT_c(\rho_{12}-\rho_{21}) -\bar{f}_L\Gamma_L\rho_{11} , \\&
\dot{\rho}_{22}=f_R\Gamma_R\rho_{00} -iT_c(\rho_{12}-\rho_{21})
-\bar{f}_R\Gamma_R\rho_{22}, \\&
\dot{\rho}_{12}=\big[-i\varepsilon -{1\over
2}(\bar{f}_L\Gamma_L+\bar{f}_R\Gamma_R)\big]\rho_{12}
+iT_c(\rho_{11}-\rho_{22}).
\end{align}
\end{subequations}
 Furthermore, in the large bias limit, $f_L=1,~f_R=0$, the
above rate equations lead to Stoof-Nazarov's rate equations
\cite{Stoof}:
\begin{subequations}
\label{i-re2}
\begin{align}
&\dot{\rho}_{00}=-\Gamma_L\rho_{00}+\Gamma_R\rho_{22},
 \\& \dot{\rho}_{11}=\Gamma_L\rho_{00}
+iT_c(\rho_{12}-\rho_{21}), \\& \dot{\rho}_{22}=-\Gamma_R\rho_{22}
-iT_c(\rho_{12}-\rho_{21}),
\\& \dot{\rho}_{12}=(-i\varepsilon -{\Gamma_R\over2})\rho_{12}
+iT_c(\rho_{11}-\rho_{22}).
\end{align}
\end{subequations}

In a summary, we have derived in this section an exact master
equation for the double dot gated by electrodes, and the
corresponding Bloch-type rate equations (\ref{ni-re}) without
considering the inter-dot Coulomb interaction as well as the rate
equations (\ref{i-re}) for the strong inter-dot Coulomb repulsion.
For convenience, we call Eq.~(\ref{ni-re}) the interaction-free
rate equations and Eq.~(\ref{i-re}) the strong-interaction rate
equations hereafter. Other approximated rate equations that have
been used in the literature are obtained at well defined limit of
the present formulae.

\section{Non-Markovian dynamics of charge qubit}
With the above formulae we can now systematically explore the
non-Markovian dynamics of the charge qubit for this double dot
system. The coherence (decoherence) dynamics of electron charges
in the double dot is determined by its internal structure as well
as external operations. The internal structure includes the
spectral properties of the reservoirs as well as the couplings
between the dots and the reservoirs embedded in the spectral
density. The external operations include charge qubit
initialization, its coherence manipulation and the qubit state
readout through the bias controls of the source and drain
electrodes. Non-Markovian decoherence effects of these internal
structure and external operations to the charge qubit are
manifested through the time-dependent coefficients in the master
equation, which is completely determined by Eq.~(\ref{EM-3}) after
the spectral density $J_{il}(\omega)$ is specified. In fact, the
time correlation functions directly tell us the length of the
correlation time which determines to what extent the
time-dependent fluctuation and memory effect become important. The
longer the correlation time is, the more memory effect acts on the
electron dynamics in the double dot and vice versa.

To be more specific, we should first specify the spectral density
$J_{il}(\omega)$ for the source and drain electrodes. Unlike the
bosonic environment where a general spectral density $J(\omega
)=\eta \omega \Big( \frac{\omega }{\omega _{c}}\Big)^{n-1} e^{-
\frac{\omega }{\omega_{c}}}$ ($\omega _{c}$ is a high frequency
cutoff and $\eta$ is a dimensionless coupling constant) was
defined and used to classify the bosonic environment as Ohmic if $
n=1$, sub-Ohmic if $0<n<1$, and super-Ohmic for $n>1$ \cite{legt},
for a fermionic environment a general spectral density should not
be a Poisson or Gaussian type distribution function because of the
fermi statistics. Here we shall use a Lorentzian spectral density
that has been used in the study of the influence of a measuring
lead on a single dot \cite{Elattari} and molecular wires coupling
to electron reservoirs \cite{Welack}. The Lorentzian spectral
density we used here has a form:
\begin{equation}
J_{il}(\omega)={\Gamma_{l}d_{l}^2/2\pi\over(\omega-
E_{i})^2+{d_{l}}^2} ,  \label{lsd}
\end{equation}
where $E_i$ is chosen to be the energy levels of the double dot,
and $l=L,R$ for $i=1,2$, respectively. There are two parameters in
$J_{il}(\omega)$ that characterize the time scales of the
reservoirs. The parameters $d_{L,R}$ describe the widths of the
Lorentzian distributions, which tell how many states in the
reservoirs around $E_{1,2}$ effectively involve in the electron
tunneling between the reservoirs and dots. Hence, the inverse
$d_{L,R}^{-1}$ characterize the time scales of the source and
drain electrodes. Another parameter is the electron tunneling
strength or the tunneling rate between the reservoirs and dots,
$\Gamma_{L,R}$, its inverse characterizes the time scale of the
electron tunneling process itself between the reservoirs and dots.
Indeed, $\Gamma_{L,R}$ also describe leakage effects of electrons
from dots to the reservoirs and vice versa. Thus a Lorentzian
spectral density well depict the time scales of non-Markovian
processes in this open quantum system.

Also, the choice of a Lorentzian spectral density makes it easy to
recover the constant spectral density which has been often used in
the literature. In fact, taking the large width limit, namely
assuming all the electron states in the reservoirs has an equal
possibility for electron tunneling between the reservoir and dot,
then $2\pi J_{il}(\omega)
\stackrel{\longrightarrow}{_{_{d_{l}\rightarrow\infty}}}\Gamma_{l}$
 reproduces the constant spectral density that has often been
used in the study of both quantum transport and quantum
decoherence phenomena in nanostructures. In this limit, the time
scale of the reservoirs is suppressed. Furthermore, in the
previous investigations, especially in the study of quantum
transport phenomena, one also takes a long time limit. Combining
these two limits (the constant spectral density and long time
limit) together, the exact master equation is reduced to the
Bloch-type rate equations in the Markovian approximation obtained
by others \cite{tb,Stoof,Gurvitz}, as we have shown in the last
section. Hence, with a Lorentzian spectral density, it is not only
convenient to analyze in details the non-Markovian dynamics but
also enables us to easily make a comparison with the Markovian
dynamics.

For a Lorentzian spectral density, the corresponding
temperature-independent time correlation function can be exactly
calculated. The result is
\begin{eqnarray}
F_{il}(\tau-\tau')= {\Gamma_l d_l \over2}\exp\big\{-(d_l +
iE_i)|\tau-\tau'|\big\} . \label{crfnc}
\end{eqnarray}
Obviously, $d_{L,R}^{-1}$ describes the correlation times of the
reservoirs. The wider/narrower it is, the shorter/longer the
correlation time will be. The internal structure of the double dot
is characterized by the energy level splitting $\varepsilon
=E_{1}-E_{2}$ and the inter-dot tunnel coupling $\Delta=2T_c$, its
time scale $T_0$ is given by the inverse of the bare Rabi
frequency $\Omega_0=\sqrt{\varepsilon^2 + \Delta^2}$.  The
non-Markovian dynamics should be dominated when the two typical
time scales, $\Omega_0^{-1}$ and $d_{L,R}^{-1}$, are in the same
order of magnitude. There is another time scale, the reservoirs'
temperature $\beta=1/kT$ that also influences the non-Markovian
dynamics of the charge qubit in certain cases. In the current
experiments for charge qubit manipulation \cite{hayashi}, the
temperature is roughly fixed at 100 mK. We will take this
temperature throughout our analysis to the charge qubit
decoherence.

\subsection{Time dependent coefficients in the master equation and non-Markovian dynamics}

Once the spectral density is specified, the full non-Markovian
dynamics of charge qubit in the double dot can be depicted using the
master equation (\ref{emaster}), or more specifically the
corresponding Bloch-type rate equations (\ref{ni-re}) and
(\ref{i-re}) for the cases of no inter-dot Coulomb repulsion and
strong inter-dot Coulomb repulsion double dot, respectively. To
solve the master equation or equivalently the rate equations, we
must determine first the time-dependent coefficients contained in
these equations, namely the shifted (renormalized) energy level
splitting $\varepsilon'(t)=E'_1(t)-E'_2(t)$ and inter-dot tunneling
coupling $\Delta'(t)=2T'_c(t)$, as well as the
dissipation-fluctuation coefficients $\Gamma(t)$ and
$\Gamma^\beta(t)$. These transport coefficients are completely
determined by the functions $u(t)$ and $v(t)$ as the solutions of
the dissipation-fluctuation equations of motion (\ref{EM-3}) which
have to be solved numerically for a given spectral density.

Using the Lorentzian spectral density (\ref{lsd}), we can
calculate explicitly all the time-dependent transport coefficients
in the master equations and then discuss the corresponding
non-Markovian dynamics by comparing with the Markovian limit in
various different time scales. We shall first analyze the
time-dependent coefficients for the charge qubit initialization
where a bias is applied to the double dot and the double dot is
adjusted to be off-resonance, i.e. $eV_{SD}=\mu_L-\mu_R \neq 0$
and $\varepsilon=E_{1}-E_{2}\ne 0$ \cite{hayashi}. We will examine
when the large bias limit is reached and how the initialization
works. After that we will go to the coherent manipulation regime
where the double dot is set up symmetrically ($E_{1}=E_{2}$) and
the chemical potentials of the electron reservoirs are aligned
above the energy levels of two dots with zero bias voltage
($\mu_L=\mu_R$). The time dependence of $\varepsilon'(t), T'_c(t)$
as well as $\Gamma(t)$ and $\Gamma^\beta(t)$ in this regime will
tell us when the non-Markovian dynamics becomes important during
the charge qubit evolution.

The dissipation-fluctuation equations of motion (\ref{EM-3}) show
that only the solution of $v(t)$ depends on the fermi distribution
function in the reservoirs. In other words, only
$\Gamma^{\beta}(t)$ sensitively depends on the bias. Other
coefficients, $\varepsilon'(t), T'_c(t)$ and $\Gamma(t)$, do not
depend on the chemical potentials $\mu_{L,R}$ and thus the bias.
Their time dependencies are completely determined by the internal
parameters of the double dot and the spectral density. In the
initialization scheme where a bias is presented and the energy
splitting of the two levels in the double dot is nonzero within
the transport window, the time-dependence of $\Gamma^{\beta}$ is
plotted in Fig.~\ref{fig3} by varying the bias voltage. We find
that the large bias limit is reached at about 100 $\mu$eV for the
given internal parameters $\varepsilon=30~\mu$eV,
$\Delta=10\mu$eV, $\Gamma_{L,R}=\Delta$ and $d_{L,R}=\Delta/2$,
with $\Omega_{0}=\sqrt{\varepsilon^{2}+\Delta^{2}} \simeq 32
\mu$eV being the bare Rabi frequency of the charge qubit. This
large bias limit is shown in Fig.~\ref{fig3} that the curves for
$eV_{SD}=90~\mu$eV are very close to the curves of
$eV_{SD}=430~\mu$eV, and the curves for $eV_{SD}=130~\mu$eV
perfectly overlap with the curves of $eV_{SD}=430~\mu$eV.
\begin{figure}[ht]
%\begin{center}
\begin{center}
\includegraphics[width=7.5cm, height=7.5cm]{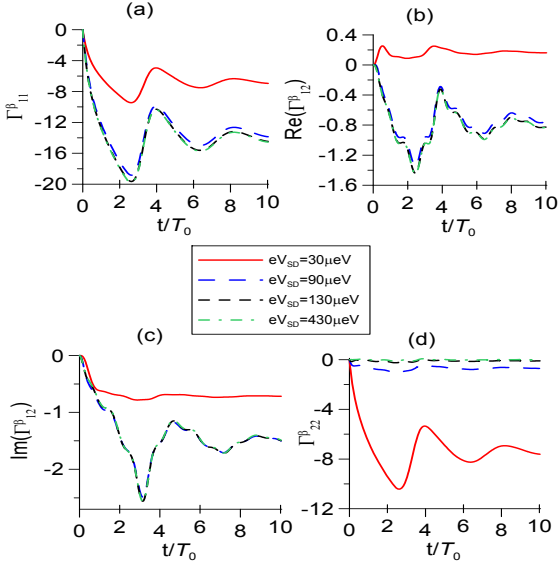}
\caption{The time-dependence of $\Gamma^{\beta}$ by varying the
bias voltage. The double dot is set to be off-resonance with
$\varepsilon=30~\mu$eV, $\Delta=10\mu$eV,  $\Gamma_{L,R}=\Delta$
and $d_{L,R}=\Omega_0/2$. (a) and (d) show the two diagonal matrix
elements of $\Gamma^{\beta}$. Since $\Gamma^{\beta}$ is hermitian
only one of the off-diagonal elements is plotted with the real
part shown in (b) and the imaginary part in (c). The period of the
bare Rabi cycle is $T_{0}=2\pi/\Omega_{0}$.} \label{fig3}
\end{center}
%\end{center}
\end{figure}

Fig.~\ref{fig4} shows the time-dependence of other coefficients for
different energy level splitting of the double dot states
($\varepsilon=0, 10, 70 \mu$eV) with different tunneling rate
($\Gamma_L=\Gamma_R=2.5, 10, 25 \mu$eV). The result shows that
$\varepsilon'(t)=0$ for $\varepsilon=0$, and the initial value
$\mbox{Re}(T_c')(0)$ always equals to $T_c$ with
$\mbox{Im}(T'_c)(0)$ being zero in all the cases we have calculated.
For a small tunneling rate ($\Gamma_{L,R} < \Delta/2$), the
time-dependence of these coefficients is almost negligible (see red
solid lines in Fig.~\ref{fig4}). The time-dependent effect appears
when the tunneling rates between the reservoirs and dots become
relatively large. These time-dependent behaviors vary sensitively on
the energy splitting of the double dot states. Increasing the energy
splitting $\varepsilon$ changes the time-dependent behaviors of all
the coefficients significantly, as shown in Fig~\ref{fig4}.
\begin{figure}[ht]
%\begin{center}
\begin{center}
\includegraphics[width=8.5cm, height=10.0cm]{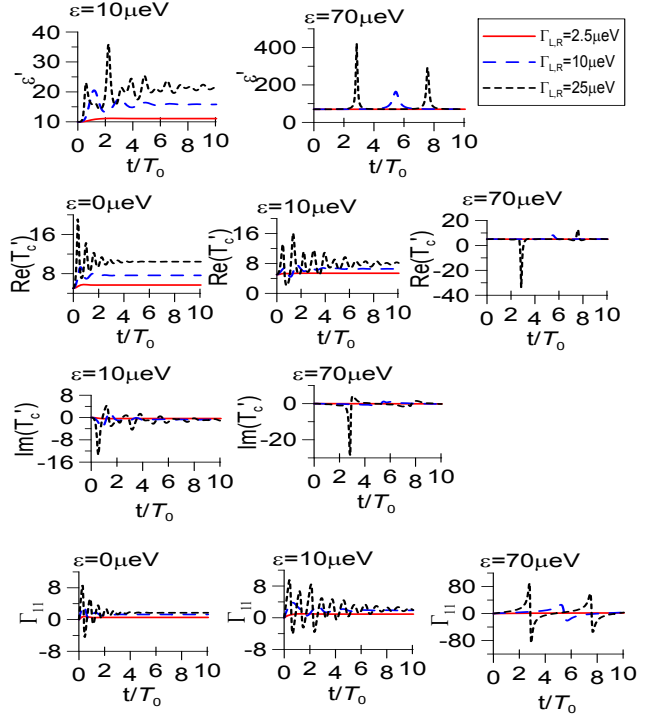}
\caption{The time-dependence of the coefficients $\varepsilon'(t),
T'_c(t)$ and $\Gamma(t)$ for different energy splitting
$\varepsilon$ and tunneling rate $\Gamma_{L,R}$ at the inter-dot
coupling $\Delta=10\mu$eV and the spectral widthes
$d_{L,R}=\Delta/2$. } \label{fig4}
\end{center}
%\end{center}
\end{figure}

The dynamics of electron charges in this initialization regime is
plotted in Fig.~\ref{fig5}. If we dope only one excess electron in
the left dot with the right dot being empty, the efficiency of
keeping this initial state decreases as the bias decreasing [see
Fig.~\ref{fig5}(a)]. If the two levels of the double dot are in
resonance ($\varepsilon=0$) or the inter-dot tunnel coupling is
larger than the tunneling rates $\Delta > 2\Gamma_{L,R}$, it also
has a low efficiency to keep the double dot in the initial state
$\rho_{11}=1$. A large bias configuration ($eV_{SD}$ larger than
100 $\mu$eV) maintains the double dot in the initial state
$\rho_{11}=1$ very well. Furthermore, for a large bias it also
quickly leads the double dot into the state $\rho_{11}\thicksim1$
even if the initial state is $\rho_{00}=1$ (both dots are empty
initially) or $\rho_{22}=1$(the left dot is empty but the right
dot is occupied by one excess electron),  as shown in
Fig.~\ref{fig5}(b). These numerical solutions are obtained using
the interaction-free rate equations (\ref{ni-re}). But the
strong-interaction rate equations (\ref{i-re}) give qualitatively
the same result for initialization.
%(where the range of bias is limited to be less than the energy of
%inter-dot Coulomb repulsion which is estimated to be 200$\mu$eV).
Taking the bias to be 650$\mu$eV and
the reservoirs' temperature to be 100mK that have used in
experiments \cite{hayashi}, a very efficient initialization of the
charge qubit can be obtained, as shown in Fig.~\ref{fig5}.
\begin{figure}[ht]
%\begin{center}
\begin{center}
\includegraphics[width=8.25cm, height=4.0cm]{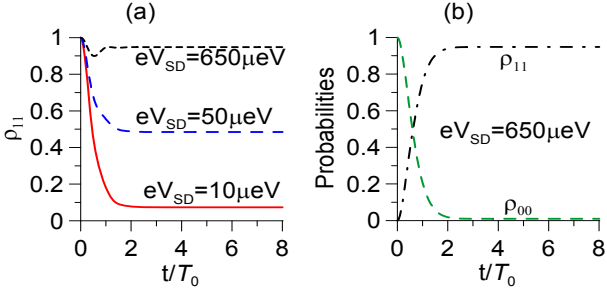}
\caption{(a) The time evolution of $\rho_{11}$, the probability of
finding one excess electron in the left dot with the right dot
being empty, is plotted at various bias when the initial state is
$\rho_{11}=1$. (b) The time evolutions of $\rho_{11}$(black
dash-dotted line) and $\rho_{00}$(green long dashed line) when the
initial state is an empty state ($\rho_{00}=1$), where $eV_{SD}=
650 \mu$eV. Other input parameters used here are the same as that
in Fig.~\ref{fig3}.} \label{fig5}
\end{center}
%\end{center}
\end{figure}

Meanwhile, Fig.~\ref{fig4} shows that for a relatively large
tunneling rate $\Gamma_{L,R}$, the smooth time oscillation of all
the coefficients at small $\varepsilon$ become discontinuing at a
relatively large $\varepsilon$ value. Such discontinuities
correspond to the electron hopping to the localized charge states
in the double dot where the inter-dot tunnel coupling $\Delta$ is
almost negligible in comparison with the level splitting
$\varepsilon$. Such discontinuities are also manifested perfectly
in the electron dynamics. The discontinuities coincide with the
times at which the electron is found in a localized charge state
of the double dot in a very high probability.
Fig.~\ref{fig6}(a)-(b) are obtained using the rate equations
(\ref{ni-re}) and (\ref{i-re}), respectively, where we plot the
corresponding electron charge dynamics together with the
time-dependence of $\Gamma_{11}$. We find that for a large bias,
the electron charge dynamics given by the interaction-free rate
equations (\ref{ni-re}) and the strong-interaction rate equations
(\ref{i-re}) display the same feature, including the coincidence
between the discontinuities in the time-dependent coefficients and
the emergence of a localized charge state in the double dot. Note
that if a sufficiently large bias is applied across the double
dot, initialization can still be achieved regardless of these
discontinuities. This is because a large bias is the most dominant
factor in this situation. As a conclusion, initialization of
charge qubit in the double dot can be easily achieved in a large
bias limit with relatively small tunnel coupling and tunneling
rates.  All these parameters are tunable in experiments.
\begin{figure}[ht]
%\begin{center}
\begin{center}
\includegraphics[width=8.cm, height=8.5cm]{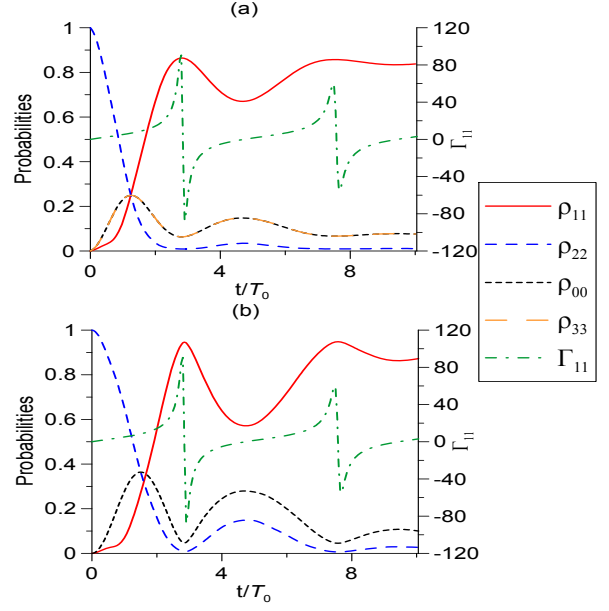}
\caption{The coincidence between the discontinuities in
$\Gamma_{11}(t)$ (the green dot-dashed line) and the emergence of
a localized charge state in the double dot. The input parameters
$\varepsilon=70 \mu$eV, $\Delta=10\mu$eV, $d_{L,R}=\Delta/2$ and
$\Gamma_{L,R}=25\mu$eV with the bias $eV_{SD}=120 \mu$eV such that
the initialization can be achieved. (a) for the double dot without
considering the inter-dot Coulomb repulsion. (b) for the double
dot in the strong inter-dot Coulomb repulsion regime.}
\label{fig6}
\end{center}
%\end{center}
\end{figure}

Now we turn into the regime for charge qubit rotations where the
double dot is set up at the resonant levels ($E_1=E_2=E$) and the
fermi surfaces of the electron reservoirs are aligned above the
resonant levels ($\mu_L=\mu_R=\mu$ and $\mu-E > 0$). In other
words, the double dot is set to be symmetric and unbiased for
charge coherence manipulation \cite{hayashi}.
Figs.~\ref{fig7}-\ref{fig9} show the time-dependencies of the
shifted inter-dot tunnel coupling $\Delta'(t)$, as well as the
dissipation-fluctuation matrices $\Gamma(t)$ and $\Gamma^\beta(t)$
for the symmetric double dot by varying the chemical potential
$\mu-E$, the spectral widths $d_{L,R}$ and the tunneling rates
$\Gamma_{L,R}$, respectively, from which we can determine the time
scales within which non-Markovian processes dominate the charge
coherence dynamics at zero bias. We find that for the symmetric
double dot, the shifted energy level splitting $\varepsilon'(t)$
and the imaginary part of the shifted inter-dot tunnel coupling
$\mbox{Im}\Delta'(t)$ keep to be zero as we have already pointed
out in Fig.~\ref{fig4}. The off-diagonal element $\Gamma_{12}(t)$
and the imaginary part of $\Gamma^{\beta}_{12}(t)$ are found also
to be zero, while the diagonal elements
$\Gamma_{11}(t)=\Gamma_{22}(t)$ and $\Gamma^{\beta}_{11}(t)
=\Gamma^{\beta}_{22}(t)$. All the time-dependent coefficients
change in time in the beginning and then approach to an asymptotic
value (the Markovian limit) at different time scales. These
time-dependence behaviors will be used to analyze the decoherence
dynamics of charge qubit in the next subsection.

In Fig.~\ref{fig7} we plot the shifted inter-dot tunnel coupling
$\Delta'(t)$, the dissipation-fluctuation matrices
$\Gamma_{11}(t)$ and $\Gamma^\beta_{11,12}(t)$ by varying the
chemical potentials with respect to the energy levels of the
double dot, $\mu-E$.  The red solid, blue long dashed and the
black short dashed lines correspond to $\mu-E=0,~25$ and 50$\mu$eV
respectively. In Fig.~\ref{fig7} the spectral widths
$d_{L,R}=\Omega_{0}/2$, where $\Omega_{0}=\Delta$ is the bare Rabi
frequency of the double dot. In other words, the time scale of the
reservoirs is chosen about two Rabi cycles of the system. A large
amplitude variation of these time-dependent coefficients within
the time scale of the reservoirs is clearly shown up in the
figure. After that time, all these time-dependent coefficients
approach to a steady value which corresponds to their asymptotic
values as a Markovian limit. This indicates that the possible
non-Markovian dynamics is mainly caused by the time-fluctuations
of these coefficients within the characteristic time of the
reservoirs.
\begin{figure}[ht]
%\begin{center}
\begin{center}
\includegraphics[width=7.0cm, height=7.0cm]{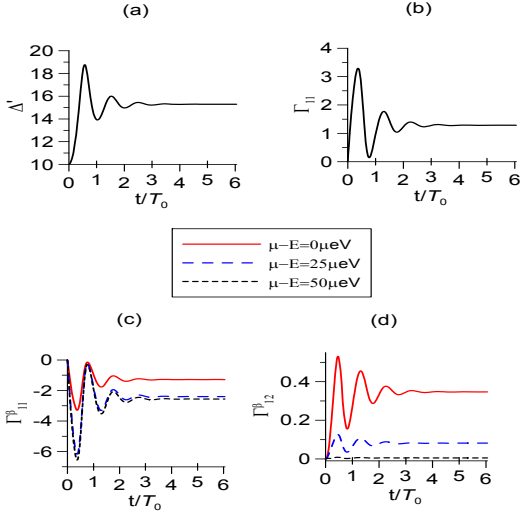}
\caption{(a) the shifted inter-dot tunnel coupling $\Delta'(t)$,
(b)-(d) the dissipation-fluctuation matrix elements
$\Gamma_{11}(t)$, and $\Gamma^\beta_{11,12}(t)$ by varying the
chemical potential $\mu=\mu_{L,R}$. The dot parameters $E=E_1=E_2$
and $\Delta=10~\mu$eV.  $\Gamma_{L,R}=\Delta$ and
$d_{L,R}=\Delta/2$. The matrix $\Gamma^{\beta}$ is hermitian so
that only one of the off-diagonal elements is presented here. The
red solid, blue long dashed and the black short dashed lines
correspond to $\mu-E=0,~25$ and 50~$\mu$eV respectively.}
\label{fig7}
\end{center}
%\end{center}
\end{figure}

Fig.~\ref{fig8} is the same plot with varying the electron
tunneling rates between the reservoirs and dots but fixing the
chemical potentials at $\mu-E=\Delta$. The small tunneling rate
$\Gamma_{L,R}=2.5~\mu$eV ($ \leq \Delta/2$) does not show a
significant time variation in the dissipation-fluctuation
coefficients, as also shown in Fig.~\ref{fig4}. The shifted
inter-dot tunnel coupling at this tunneling rate is very close to
the bare one ( $\sim \Delta$), see the red solid line in
Fig.~\ref{fig8}(a). This implies that a small electron tunneling
rate between the reservoirs and dots (corresponds to a small
leakage effect) does not manifest the non-Markovian dynamics
significantly. Increasing the tunneling rates enlarges the charge
leakage effect from the reservoirs to dots and vice versa, thus
enhances the non-Markovian effects as well, as shown by the
giggling and wiggling time evolutions of the
dissipation-fluctuation coefficients in the figure. The
time-dependence of shifted inter-dot tunnel coupling $\Delta'$ at
large tunneling rates also has a significant shift from the bare
one $\Delta$ besides the oscillation within the non-Markovian time
region. Meanwhile, the charge oscillation frequency has different
shifts from the bare Rabi frequency for different $\Gamma_{L,R}$
values. How these time-dependent (non-Markovian) behaviors
influent the charge coherence will be discussed in details in the
next subsection.
\begin{figure}[ht]
%\begin{center}
\begin{center}
\includegraphics[width=7.25cm, height=6.5cm]{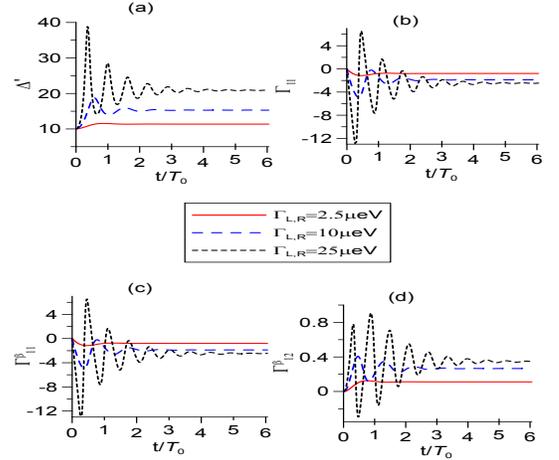}
\caption{The same plot as in Fig.~\ref{fig7} with fixing the
chemical potentials at $\mu-E=\Delta$ and varying the electron
tunneling rates between the reservoirs and dots $\Gamma_{L,R}$
symmetrically. The other parameters are the same as in
Fig.~\ref{fig7}.} \label{fig8}
\end{center}
%\end{center}
\end{figure}

Fig.~\ref{fig9} shows how the time-dependence of the shifted
inter-dot tunnel coupling and the dissipation-fluctuation
coefficients change by varying the spectral widths $d_{L,R}$. We
plot these time-dependent coefficients for three different
spectral widths: $d_{L,R}$ = 1, 5 and 25 $\mu$eV. The result shows
that when $ d_{L,R}=25~\mu$eV, the dynamics of the double dot
already reaches to the Markovian limit, namely all the
time-dependent coefficients approach to their asymptotic values in
a very short time (less than a half cycle of the bare Rabi
oscillation). When the spectral widths $d_{L,R}$ becomes small
($\leq \Delta$) so that the characteristic time of the reservoirs
becomes long), the time oscillation of all the transport
coefficients becomes strong. Correspondingly the charge dynamics
is dominated by non-Markovian processes.
\begin{figure}[ht]
%\begin{center}
\begin{center}
\includegraphics[width=7.25cm, height=6.5cm]{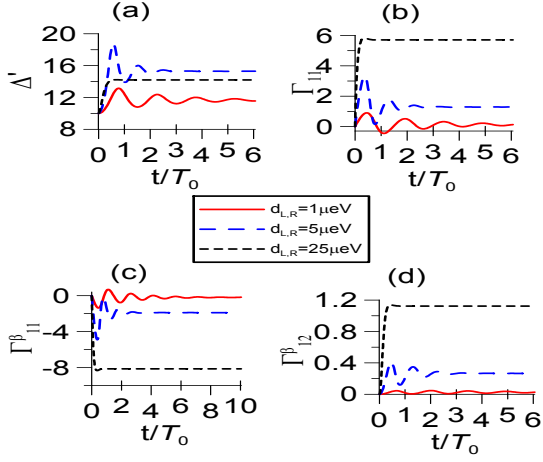}
\caption{The same plot as in Fig.~\ref{fig8} by varying $d_{L,R}$
but fixing $\Gamma_{L,R}=\Delta$ and other parameters are those
used in Fig.~\ref{fig8}.} \label{fig9}
\end{center}
%\end{center}
\end{figure}

The above numerical results tell that the spectral widths
$d_{L,R}$ of the reservoirs (mainly as a memory effect) and the
tunneling rates $\Gamma_{L,R}$ between the reservoirs and dots
(mainly as a leakage effect) are two basic parameters to
characterize the occurrence of non-Markovian dynamics in this
double dot device. Comparing the results of Figs.~\ref{fig7},
\ref{fig8} and \ref{fig9}, we find that for the coherence
manipulation of charge qubit where the the double dot is unbiased
\cite{hayashi}, the time for the dissipation-fluctuation
coefficients reaching to a steady limit depends on the spectral
widths of the tunneling spectra as well as the tunneling rates
between the reservoirs and dots. The Markovian limit often used in
the literature is valid for the electron reservoirs having a
relatively small tunneling rate (negligible leakage effect) and a
large spectral width (negligible memory effect). The former
implies the validity of the Born approximation and the latter
corresponds to the Markovian approximation. The chemical
potentials of the electron reservoirs controled by the external
bias voltage just modifies the values of the
dissipation-fluctuation coefficients $\Gamma^{\beta}$ without
altering the characteristic times of the reservoirs and the
system, as shown in Figs.~\ref{fig3} and \ref{fig7}. However, the
chemical potential can be very efficient to suppress the leakage
effect. Thus, $\mu-E$ is a competitive control parameter in the
coherence control of charge qubit, as we will see later.

In order to show clearly when non-Markovian or Markovian processes
play a major role in charge qubit decoherence, we take
$\Gamma_{11}$ as an example to examine at what time this
dissipation-fluctuation coefficient reaches its steady value by
varying $d_{L,R}$ and $\Gamma_{L,R}$. The result is plotted in
Fig.~\ref{fig10}. The lines signify transition times between the
time-dependent fluctuating and the steady dissipation-fluctuation
coefficients by varying tunneling rate at a given spectral width.
Non-Markovian dynamics can be seen mostly in the time range under
the lines. It shows that the strong non-Markovian dynamics
corresponds to a relatively small spectral widths $d_{L,R}$ (a
strong memory effect) and a relatively large tunneling rates
$\Gamma_{L,R}$ (a large leakage effect) compared with the
inter-dot tunnel coupling $\Delta$. Non-Markovian dynamics
disappears for a large $d_{L,R}$ ($ \geq 2\Delta$) and a small
$\Gamma_{L,R}$ ($ \leq \Delta/2$). In the parameter range of
interests to the experiments\cite{prs}, the non-Markovian
processes do not go over more than five Rabi cycles, with which a
significant effect can be seen in maintaining charge coherence, as
we will see below.
\begin{figure}[ht]
%\begin{center}
\begin{center}
\includegraphics[width=7.25cm, height=6.5cm]{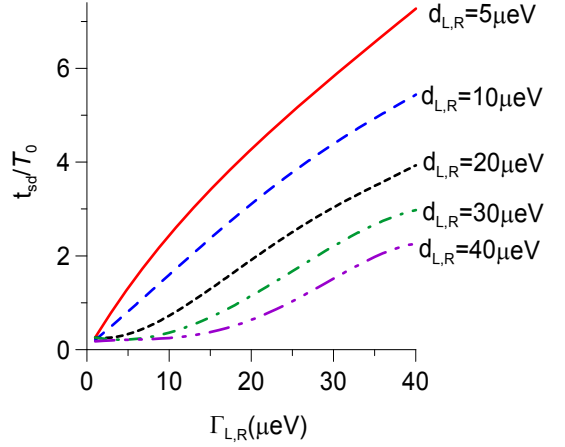}
\caption{The lines to separate the Markovian and the non-Markovian
dynamics in the $\Gamma_{11}-t$ plot, each line corresponds to a
given value of $d_{L,R}$. The parameters $\mu-E=\Delta=10 \mu$eV
are used.} \label{fig10}
\end{center}
%\end{center}
\end{figure}

\subsection{Decoherence dynamics of charge qubit}

Have examined the time dependencies of all the transport
coefficients (the shifted energy level splitting
$\varepsilon'(t)$, the renormalized inter-dot tunnel coupling
$\Delta'(t)$ and the dissipation-fluctuation coefficients
$\Gamma(t), \Gamma^\beta(t)$) in the master equation for both
biased and unbiased double dot, we shall discuss now the
decoherent dynamics of the charge qubit in this subsection.
Experimentally the coherence manipulation of charge qubit is
performed for the double dot on resonance,
$\varepsilon=E_1-E_2=0$. The corresponding shifted energy level
splitting $\varepsilon'(t)$ remains zero. Then the energy
eigenbasis of the charge qubit refers actually to the molecular
anti-bonding and bonding states, namely,
$|\pm\rangle\equiv{1\over\sqrt{2}}(|1\rangle\pm|2\rangle)$. The
oscillation between the coherently coupled localized charge states
$|1\rangle$ and $|2\rangle$ as coherent superpositions of the
molecular states describes the charge coherence, where the
renormalized Rabi frequency $\Omega(t)=\Delta'(t)$ is just the
shifted inter-dot tunnel coupling for the symmetric double dot.
The time-dependent dissipation-fluctuation coefficients
$\Gamma(t), \Gamma^\beta(t)$ will disturb this coherent
oscillation and cause charge qubit decoherence.

To be specific, we let the initial state be $\rho_{11}=1$ and
examine the time evolution of the density matrix under various
conditions. We calculate first the rate equations (\ref{ni-re}) for
the no-Coulomb-interacting double dot. The typical population
evolution shown in Fig.~\ref{fig11} tells that the double occupancy
is favored. In fact, the charge qubit of a double dot is designed in
the strong inter-dot Coulomb blockade regime where the state of
simultaneous occupation of two dots is excluded. In other words,
unlike the charge qubit initialization where both the
interaction-free and the strong interaction rate equations give
qualitatively the same result, for the coherence control of the
charge qubit the interaction-free rate equations (\ref{ni-re}) is
invalid.  The charge qubit dynamics must be described by the
strong-interaction rate equations (\ref{i-re}).
\begin{figure}[ht]
\begin{center}
\begin{center}
\includegraphics[width=5.5cm, height=5.0cm]{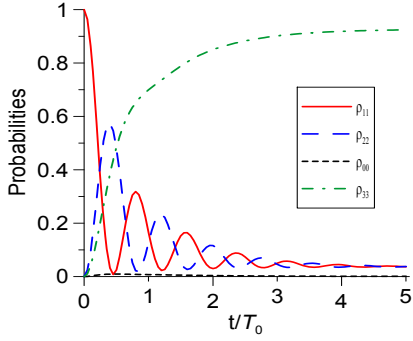}
\caption{The time evolutions of the populations $\rho_{11}$(the
red solid line), $\rho_{22}$(the blue long-dashed line),
$\rho_{00}$(the black short dashed line), and $\rho_{33}$(the
green dashed-dotted line) calculated using the interaction-free
rate equations (\ref{ni-re}) in the symmetric double dot with
 $\Delta=10 \mu$eV, $\mu-E=30\mu$eV,
$d_{L,R}=\Delta/2$ and $\Gamma_{L,R}=d_{L,R}$.  Without
considering the inter-dot Coulomb repulsion, raising up of the
chemical potentials $\mu>E$ accumulates charges into the double
dot.} \label{fig11}
\end{center}
\end{center}
\end{figure}
For the rate equations (\ref{i-re}) in the strong inter-dot
Coulomb repulsion regime to be held for charge qubit manipulation,
the energy difference between the fermi surfaces of the reservoirs
and the energy levels of the dots, $\mu-E$, cannot be too large.
The inter-dot Coulomb repulsion in the samples is estimated to be
200$\mu$eV \cite{hayashi}. The value of $\mu-E$ that can be taken
most safely should be not larger than $50\mu$eV, a quarter of the
inter-dot Coulomb repulsion energy. If $\mu-E$ is taken over 100
$\mu$eV, it is comparable to the Coulomb repulsion energy so that
the doubly occupied state cannot be completely excluded. For
convenience and consistency with the discussion in the previous
subsection, we still take the quantum dot parameters
$E_{1}=E_{2}=E$ and $\Delta=10\mu$eV. As one will see with
$\mu-E=25 \sim 50 \mu$eV, and $\Gamma_{L,R}$ and $d_{L,R}$ being
in a reasonable range, the charge qubit can maintain coherence
very well.

In Fig.~\ref{fig12}, we plot the time evolution of the reduced
density matrix by varying the aligned fermi surfaces.  The result
shows that the coherent oscillation of the charge qubit depends
sensitively on the hight of aligned fermi surfaces from the
resonant levels of the double dot, i.e., $\mu-E$. When $\mu-E$ is
not too large ($ < \Delta$) and $\Gamma_{L,R}$ is not too small
($> \Delta$), the population $\rho_{11}$ decays very quickly. This
is because although the state of simultaneous occupation of two
dots is excluded, there is still chance for electrons to escape
from the dots into the reservoirs such that both dots become
empty, namely $\rho_{00} \ne 0$ [see Fig.~\ref{fig12}(b)] as a
leakage effect. This effect can be suppressed when the fermi
surfaces are aligned such that $\mu-E$ must be relatively larger
than $\Gamma_{L,R}$. As a result, the charge qubit can maintain
the coherence very well. In Fig.~\ref{fig12}(a), we
 see that the charge coherence is perfectly maintained for
$\mu-E=50 \mu$eV. Meanwhile, the real part and the imaginary part
of the off-diagonal density matrix element $\rho_{12}$ exhibit
quite different dynamics. The imaginary part of $\rho_{12}$ has a
similar oscillatory feature as $\rho_{11}$ [see
Fig.~\ref{fig12}(d)], which depicts the coherent tunnel coupling
between two dots. While the real part of $\rho_{12}$ goes down to
be negative [see Fig.~\ref{fig12}(c)] which is related to the loss
of the energy (dissipation). We find that to maintain a good
coherent dynamics for charge qubit, the fermi surfaces of the
reservoirs is better to be aligned above the energy levels of the
double dot not less than $2\Delta$.
\begin{figure}[ht]
%\begin{center}
\begin{center}
\includegraphics[width=8.5cm, height=8.0cm]{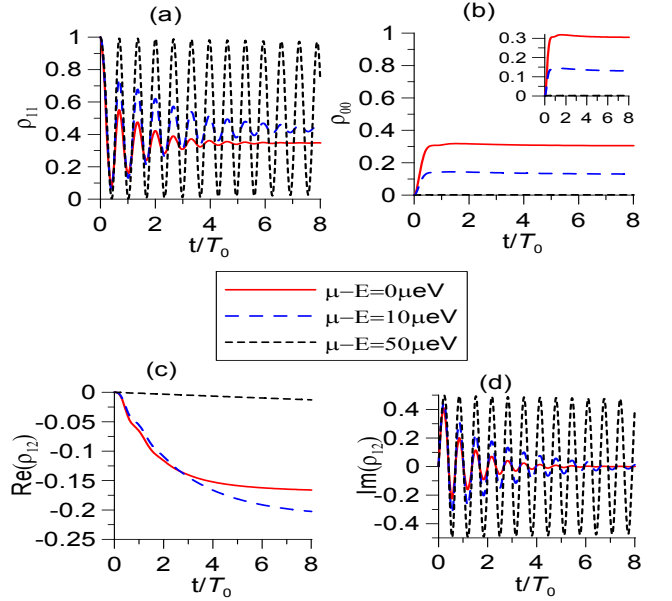}
\caption{The time evolutions of the density matrix at different
$\mu-E$ with $E$ fixed.  The red solid line is for $\mu-E=0$, the
blue long-dashed line is for $\mu-E=10\mu$eV and the black
short-dashed line is for $\mu-E=50\mu$eV. $d_{L,R}=\Delta/2$ and
$\Gamma_{L,R}=\Delta$.} \label{fig12}
\end{center}
%\end{center}
\end{figure}

In Fig.~\ref{fig13} we plot the time evolution of the reduced
density matrix at a few different tunneling rates but fixing other
parameters. Experimentally, the tunneling rates between the dots
and the reservoirs are also tunable. We fixed the spectral widthes
at $d_{L,R}=25\mu$eV for which the memory effect is largely
suppressed. Meantime, we take $\mu-E=50\mu$eV (much larger than
$\Gamma_{L,R}$) such that the leakage effect is also largely
suppressed. When the tunneling rates are small ($< \Delta/2$), the
time-dependence of these transport coefficients in the master
equation are negligible (as shown in Fig.~\ref{fig8}) so that no
non-Markovian dynamics can be observed. The corresponding charge
dynamics is given by the red solid lines in Fig.~\ref{fig13} where
the oscillation frequency is time independence, consistent with
the result in Markovian approximation. When the tunneling rates
become large ($\geq \Delta$), the charge frequency is largely
shifted and varies in time. The larger the tunneling rates are,
the faster the electron oscillates between two dots and the
reservoirs, thus the stronger the non-Markovian dynamics occurs.
Here the decay of the coherent charge oscillation (see the blue
long-dashed and black short-dashed lines in Fig.~\ref{fig13}) is
not due to the charge leakage (which has been mainly suppressed by
raising up the fermi surfaces) but a back-action decoherence
effect of the reservoirs when the tunneling rate becomes larger.
Increasing the tunneling rates leads to more charge leakage.
However, comparing the magnitude of $\rho_{00}$ with that of
$\rho_{11}$ shows that the damping of coherent charge oscillation
in the presence of a large tunneling rate is not primarily due to
charge leakage but a non-Markovian back-action decoherence effect.
\begin{figure}[ht]
%\begin{center}
\begin{center}
\includegraphics[width=8.5cm, height=8.5cm]{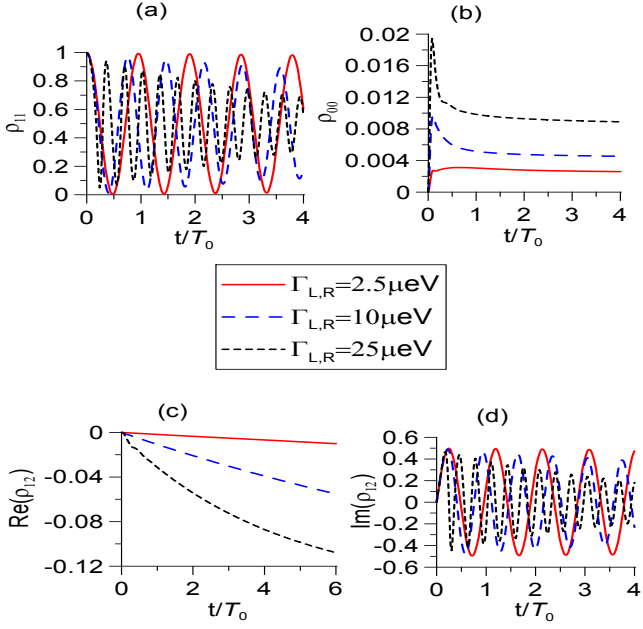}
\caption{The time evolutions of the density matrix at different
coupling constants.  The red solid line is for
$\Gamma_{L,R}=2.5\mu$eV, the blue long-dashed line is for
$\Gamma_{L,R}=10\mu$eV and the black short-dashed line is for
$\Gamma_{L,R}=25\mu$eV. The chemical potentials are kept at
$\mu-E=50\mu$eV and $d_{L,R}=25\mu$eV.} \label{fig13}
\end{center}
%\end{center}
\end{figure}

The more non-Markovian dynamics can be seen by varying the
spectral widths $d_{L,R}$. When the spectral width is comparable
to the inter-dot coupling $\Delta$, namely the characteristic time
of reservoirs is comparable to the characteristic time of the
double dot, the non-Markovian dynamics becomes the most
significant in the time evolution of charge coherence. Although it
is currently not clear how to tune the spectral widths in
experiments, it is still interesting to see what roles the
spectral widths (or more generally speaking, a non-constant
spectral density) play in the charge decoherence dynamics. We plot
in Fig.~\ref{fig14} the time evolutions of the density matrix
elements at various spectral widths with $\mu-E=50\mu$eV and
$\Gamma_{L,R}=\Delta$. As we can see if $d_{L,R}$ is small ($\leq
\Delta/2$), the coherent charge dynamics is well preserved
although it is a strong non-Markovian process. With increasing the
spectral widths, the decoherent charge dynamics becomes visible
and also becomes Markov type. Widening the spectral widthes damps
the coherent charge oscillation. Wider spectral width also causes
more charge leakage. But comparing the magnitudes of $\rho_{00}$
with $\rho_{11}$ in Fig.~\ref{fig14} shows again that the damping
of coherent charge oscillation in the presence of a wide spectral
width is a Markovian back-action decoherence effect. The
frequencies of coherent charge oscillation are also shifted
differently from the bare Rabi frequency for different spectral
widths presented. [see Fig.~\ref{fig14}(a)-(d)].
\begin{figure}[ht]
%\begin{center}
\begin{center}
\includegraphics[width=8.5cm, height=8.5cm]{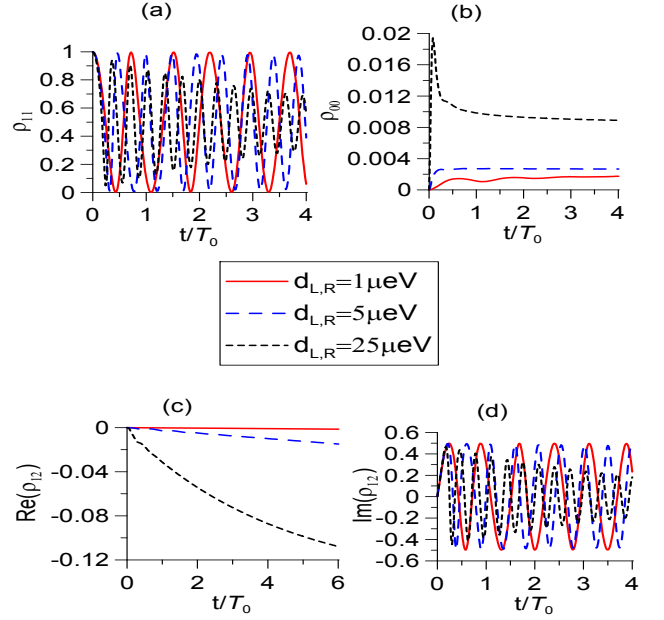}
\caption{The time evolutions of the density matrix at different
spectral widths. The red solid line is for $d_{L,R}=1\mu$eV, the
blue long-dashed line is for $d_{L,R}=5\mu$eV and the black
short-dashed line is for $d_{L,R}=25\mu$eV. $\mu-E$ is kept at
$50\mu$eV and $\Gamma_{L,R}=\Delta$. } \label{fig14}
\end{center}
%\end{center}
\end{figure}

From the above analysis, we find that the charge qubit coherence
can be maintained very well when either the spectral widths
$d_{L,R}$ or the tunneling rates $\Gamma_{L,R}$ are sufficiently
smaller than $\mu-E$. However, when both the spectral widths
$d_{L,R}$ and the tunneling rates $\Gamma_{L,R}$ become comparable
to $\mu-E$, the decay of charge coherency can be seen within  a
few cycles of the bare Rabi oscillation. The difference between
the non-Markovian and the Markovian process manifests in the
difference between the renormalized Rabi frequency $\Omega(t)=
\Delta'(t)$ (for the symmetric double dot) and the bare Rabi
frequency $\Omega_0=\Delta$. In Fig.~\ref{fig15}, we plot the
average renormalized Rabi freuency $\langle \Omega \rangle =
\overline{\Omega(t)}$ by varying the spectral widths $d_{L,R}$ and
the tunneling rate $\Gamma_{L,R}$, respectively, and make a
comparison with the bare Rabi frequency (the black short dashed
lines). It shows that the renormalized Rabi frequency has a large
shift from the bare one, except for the region (the Markovian
regime) with the spectral width $ d_{L,R} > 2 \Delta$ and the
tunneling rate $\Gamma_{L,R} < \Delta/2 $, where the renormalized
Rabi frequency is close to the bare one. In Fig.~\ref{fig16}, we
plot the time evolution of density matrix $\rho_{11}$ using the
strong-interaction rate equations (\ref{i-re}) for a few different
sets of ($\Gamma_{L,R}, d_{L,R}$) and make a comparison with the
Markovian approximation determined by the rate equations
(\ref{i-re1}), where the charge coherence dynamics in the double
dot from Markovian to the non-Markovian processes is clearly
demonstrated.
\begin{figure}[ht]
%\begin{center}
\begin{center}
\includegraphics[width=8.5cm, height=4.0cm]{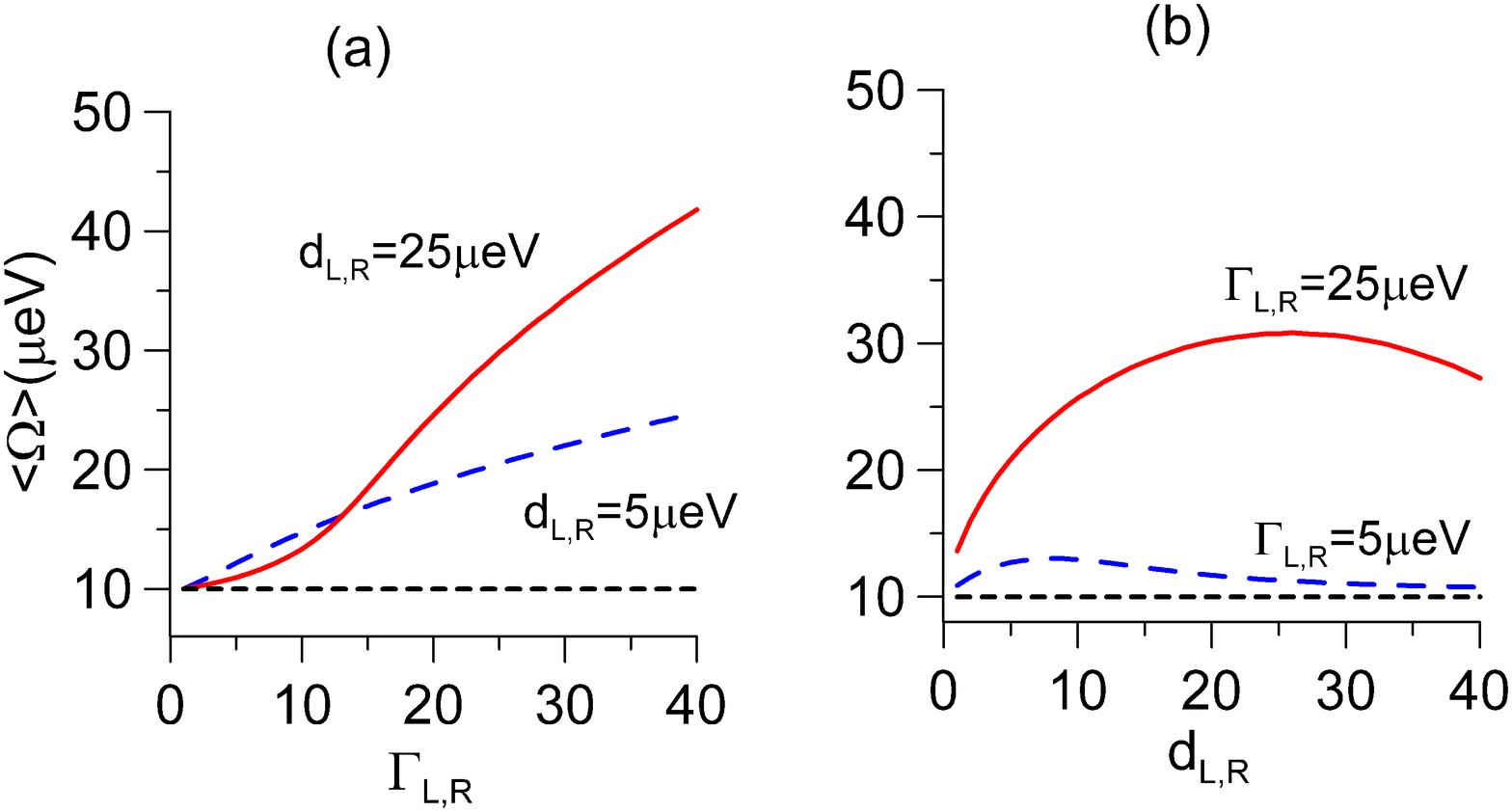}
\caption{The average renormalized Rabi frequency $\langle \Omega
\rangle = \overline{\Omega(t)}$ by varying the spectral widths
$d_{L,R}$ and the tunneling rate $\Gamma_{L,R}$, respectively, and
comparing with the bare Rabi frequency (the black short dashed
lines). Here we take $\mu-E=50\mu$eV to suppress the charge
leakage.} \label{fig15}
\end{center}
%\end{center}
\end{figure}\begin{figure}[ht]
%\begin{center}
\begin{center}
\includegraphics[width=8.5cm, height=8.5cm]{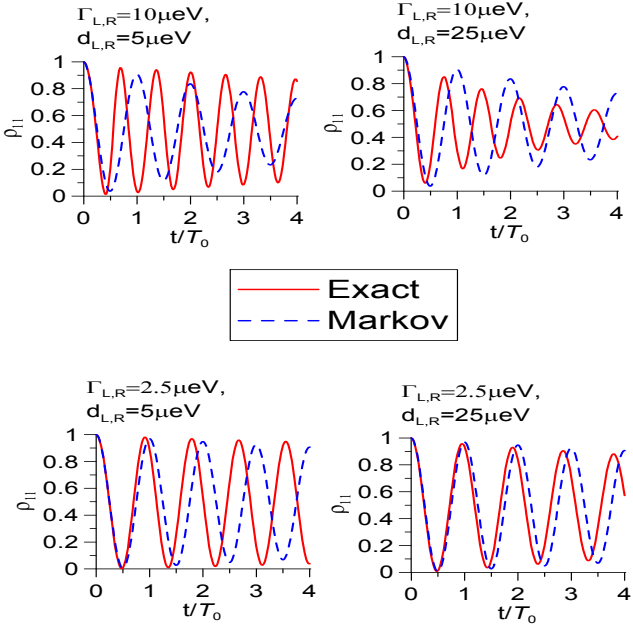}
\caption{The time evolutions of $\rho_{11}$ obtained from the
 rate equation (\ref{i-re}) and the corresponding Markovian
limit (\ref{i-re1}) for the strong-Coulomb-interacting double dot.
(a) the strongest non-Markovian regime for a small spectral width
and a large tunneling rate. (b) the non-Markovian dynamics
controlled by a large tunneling rate alone. (c) the non-Markovian
dynamics controlled by  a small spectral width alone. (d) The
Markovian regime corresponding to a small tunneling rate and large
spectral width. Here we take $\mu-E=30\mu$eV.} \label{fig16}
\end{center}
%\end{center}
\end{figure}

\subsection{Relaxation time $T_1$ and dephasing time $T_2$}
To understand quantitatively the decoherence dynamics of charge
qubit in the double dot, we shall now extract the decay rates (the
relaxation time $T_1$ and the dephasing time $T_2$) for various
manipulation conditions. The relaxation time $T_1$ characterizes
the time going from the anti-bonding state (with higher energy) to
the bonding state (with lower energy) in the energy eigenbasis
$|\pm \rangle={1\over\sqrt{2}}(|1\rangle\pm|2\rangle)$ (for
symmetric double dot). It is described by the decay of the
diagonal element of the density matrix in the energy eigenbasis,
$\langle+|\rho|+\rangle
={1\over2}(\rho_{11}+\rho_{22})+\mbox{Re}(\rho_{12})$. In the
ideal case (such as in NMR), $\rho_{11}+\rho_{22}=1$ so that $T_1$
is completely determined by the real part of $\rho_{12}$. For the
charge qubit in double dots, in general $\rho_{11}+\rho_{22}<1$
because of charge leakage. But in the practical manipulation, the
fermi surfaces of the reservoirs are set high enough from the
resonant energy levels of the double dot such that charge leakage
can be suppressed. Thus we can still extract $T_1$ from
$\mbox{Re}\rho_{12}$. The decoherence (dephasing) time $T_{2}$
corresponds to the decay of the off-diagonal element of the
reduced density matrix $\langle+|\rho|-\rangle
={1\over2}(\rho_{11}-\rho_{22})-i\mbox{Im}(\rho_{12})$. As we will
see later the time-dependencies of $\rho_{11}, \rho_{22}$ and the
imaginary part of $\rho_{12}$ behave very similar. This may tell
us that $T_2$ can be extracted either from $\rho_{11}, \rho_{22}$
or $\mbox{Im}\rho_{12}$. Experimentally, one extracted $T_2$ from
$\rho_{22}$ by measuring the current proportion to $\rho_{22}$
\cite{hayashi}.

Having made the above analysis, we shall extract the decay rates
of the charge coherent oscillations from the time evolution of the
reduced density matrix elements $\mbox{Re}\rho_{12}(t)$ and
$\mbox{Im}\rho_{12}(t)$ or $\rho_{11}(t)$ by fitting a decay
oscillating function plus an offset to the numerical data. An
intuitive fitting function for the charge oscillation decay would
be $Ae^{-Bt}\cos(\langle\Omega\rangle t)+C$.  However this fitting
function fails for $\rho_{11}(t)$. The typical damping oscillation
of $\rho_{11}(t)$ shows that it converges to a steady value of
$\rho_{11}\leq {1\over2}$ at large $t$. Thus $\rho_{11}(t)$ can be
well described by the fitting function
\begin{align}
\rho_{11}(t)=f_h(t)\cos^{2}(\langle\Omega\rangle
t/2)+f_l(t)\sin^{2}(\langle\Omega\rangle t/2) \label{ffr11}
\end{align}
where $f_{h,l}(t)= A_{h,l}\exp(-B_{h,l}t^{s})+C_{h,l}$ are used to
fit the downward shift of the peaks and the upward shift of the
valleys in the damped oscillation, respectively. The oscillating
function is the squares of sine and cosine functions with half the
oscillation frequency, $\sin^{2}(\langle\Omega\rangle t/2)$ and
$\cos^{2}(\langle\Omega\rangle t/2)$, rather than
$\cos(\langle\Omega\rangle t)$ and $\sin(\langle\Omega\rangle t)$.
This can be easily understood by considering an ideal qubit. Its
Rabi oscillation conditioned to the initial state $\rho_{11}(0)=1$
is given by
$\rho(t)=\begin{pmatrix}\cos^{2}(\Omega t/2)&-{i\over2}\sin(\Omega t)\\
{i\over2}\sin(\Omega t)&\sin^{2}(\Omega t/2)\end{pmatrix}$ in the
localized charge state basis. Similarly, the off-diagonal density
matrix elements can be described by the fitting functions
\begin{subequations} \label{ffr12} \begin{align}
&\mbox{Im}\rho_{12}(t)=\big\{A_i\exp(-B_it^{s})+C_i \big\}
   \sin(\langle\Omega\rangle t), \label{ffr12i} \\
&\mbox{Re}\rho_{12}(t)=A_r\exp(-B_rt^{s})+C_r . \label{ffr12r}
\end{align}
\end{subequations}
The fitting parameters $A_x,B_x,C_x$ for $x=h,l,i,r$ are generally
different for different fitting function and different data.
Fig.~\ref{fig17} is a plot of using the fitting function of
(\ref{ffr11}) to fit the exact numerical solution of
$\rho_{11}(t)$. The results show that with the decay function
$f_h(t)=0.48\exp{(-0.58t)}+0.50$  and
$f_l(t)=-0.47\exp(-0.54t)+0.49$, the fitting function
(\ref{ffr11}) gives very much the same solution as that obtained
numerically from Eq.~(\ref{i-re}) for $\rho_{11}(t)$.
\begin{figure}[ht]
%\begin{center}
\begin{center}
\includegraphics[width=6.5cm, height=4.5cm]{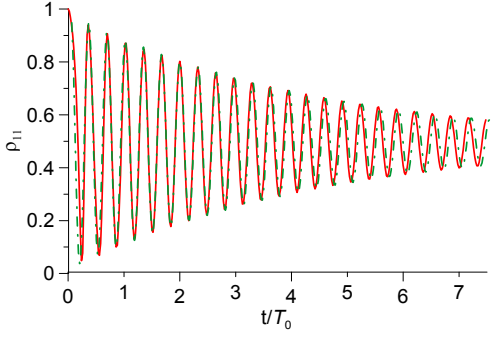}
\caption{(a) Using the fitting function
$f_h(t)\cos^{2}(\langle\Omega\rangle
t/2)+f_l(t)\sin^{2}(\langle\Omega\rangle t/2)$ (the red solid
line) to fit the exact numerical solution of $\rho_{11}(t)$ (the
green dashed-dotted line).  Here we use the parameters
$\mu-E=50\mu$eV, $\Gamma_{L,R}=25 \mu$eV and $d_{L,R}=25\mu$eV. }
 \label{fig17}
\end{center}
%\end{center}
\end{figure}

In order to have a better fitting to various tunable parameters,
we plot in Fig.~\ref{fig18} the fitting errors for $\rho_{11}(t)$
by averaging the deviations between $f_{h}(t)$ and the peaks of
the oscillating $\rho_{11}(t)$ in a large range of the chemical
potential, the tunneling rate as well as the spectral width. The
top two plots in Figs.~\ref{fig18} are the fitting errors by
varying $\mu_{L,R}$ but fixing $(\Gamma_{L,R}, d_{L,R})= (25, 5)
\mu$eV (the strong non-Markovian regime) and (5, 25) $\mu$eV (the
Markovian limit regime). We find that in the strong non-Markovian
regime, when the fermi surfaces are not far away from the resonant
levels of the double dot ($\mu-E < 25 \mu$eV), the best fitting is
a sub-exponential decay with $s< 1$. When $\mu-E
> 25 \mu$eV, the fitting function becomes a simple exponential
decay function ($s=1$), while, in the Markovian limit, the best
fitting is just a simple exponential decay for all the values of
$\mu-E$. The middle two plots in Figs.~\ref{fig18} are the fitting
errors by varying $\Gamma_{L,R}$ but fixing $(\mu-E, d_{L,R})=
(10, 5) \mu$eV (the non-Markovian regime) and (30, 25) $\mu$eV
(the Markovian limit). Again we see that in the non-Markovian
regime, the best fitting is a sub-exponential decay with $s<1$ for
all the values of $\Gamma_{L,R}$ except for some very small
$\Gamma_{L,R}$ ($< \Delta/2$) which indeed enters the Markovian
limit where the fitting function becomes a simple exponential
decay function. In the Markovian limit (the large spectral width
limit here), the best fitting is given by a simple exponential
decay ($s=1$) again. The bottom two plots in Figs.~\ref{fig18} are
the fitting errors by varying $d_{L,R}$ but fixed $(\mu-E,
\Gamma_{L,R})= (10, 25) \mu$eV (the non-Markovian regime) and (30,
5) $\mu$eV (the Markovian limit). It tells that in the
non-Markovian regime, when $d_{L,R} < \Delta$ (the strong
non-Markovian regime), the best fitting is still given by a
sub-exponential decay ($s <1$), while for $d_{L,R}
> \Delta$ the system transits to the Markovian, the fitting
function becomes again a simple exponential decay function. In the
Markovian limit (small tunneling rate limit), the best fitting
 is just given by a simple exponential decay ($s=1$), as one expected.
\begin{figure}[ht]
%\begin{center}
\begin{center}
\includegraphics[width=8.5cm, height=12.5cm]{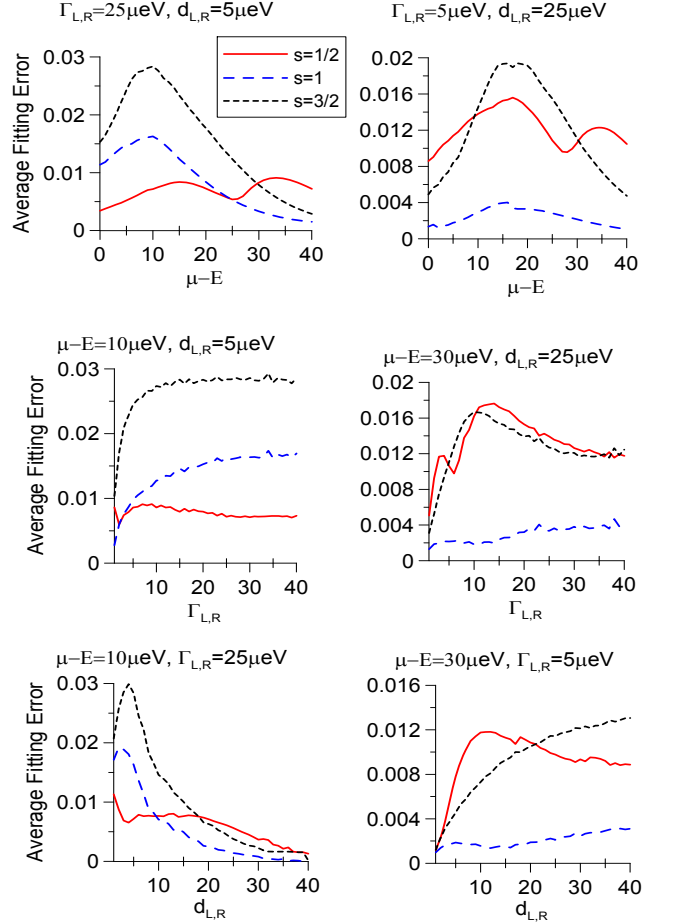}
\caption{The average fitting errors by fitting the exact numerical
solutions of $\rho_{11}(t)$ with the fitting function given by
Eq.~(\ref{ffr11}). The top two plots are the fitting errors by
varying $\mu-E$ but fixing $(\Gamma_{L,R}, d_{L,R})= (25, 5)
\mu$eV and (5, 25) $\mu$eV, respectively; The middle two plots for
varying $\Gamma_{L,R}$ but fixing $(\mu-E, d_{L,R})= (10, 5)
\mu$eV and (30, 25) $\mu$eV; and the bottom two plots are obtained
by varying $d_{L,R}$ but fixed $(\mu-E, \Gamma_{L,R})= (10, 25)
\mu$eV and (30, 5) $\mu$eV. The parameters are chosen such that
the left three figures correspond to the non-Markovian regime
while the right three figures are the Markovian limit. $s=1/2$
corresponds to the sub-exponential fitting (red lines), $s=1$ is
the simple exponential fitting (blue long-dashed lines), and
$s=3/2$ is the super-exponential fitting (black dashed lines).}
\label{fig18}
\end{center}
%\end{center}
\end{figure}

The above analysis shows that Markovian decoherence processes lead
to an exponential decay and a sub-exponential decay seems to occur
mainly in strong non-Markovian processes. But this does not imply
a simple exponential decay being necessarily Markovian. In
Fig.~(\ref{fig19}), we plot the average fitting errors of the
fitting functions (\ref{ffr12}) with the exact numerical solution
of the real and imaginary parts of the off-diagonal density matrix
element $\rho_{12}(t)$ in the strong non-Markovian regime with the
spectral widths $d_{L,R}=5 \mu$eV and the tunneling rates
$\Gamma_{L,R}=25 \mu$eV. The results show that for both
$\mbox{Im}\rho_{12}(t)$ and $\mbox{Re}\rho_{12}(t)$, the best
fitting is given by simple exponential decay for the whole range
of chemical potential up to $\mu-E = 40 \mu$eV. This forces us to
carefully look at the results in Fig.~\ref{fig18}. We find that
all the results with a sub-exponential decay show up for small
$\mu-E$ values ($<2\Delta$) where the charge leakage effect cannot
be effectively suppressed by the chemical potentials of the
reservoirs. This tells that the sub-exponential decay in
$\rho_{11}(t)$ is a charge leakage effect. The slightly different
decay behaviors for $\rho_{11}(t)$ and $\mbox{Im}\rho_{12}(t)$
actually come from the charge leakage effect contained in
$\rho_{11}(t)$ rather than a consequence of non-Markovian
dynamics, as we will see more later.
\begin{figure}[ht]
%\begin{center}
\begin{center}
\includegraphics[width=8.5cm, height=4.5cm]{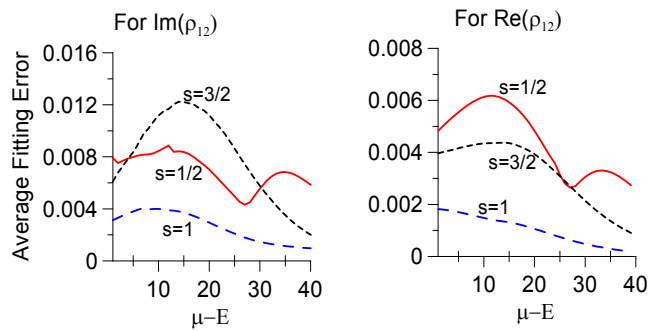}
\caption{The average fitting errors by fitting the exact numerical
solution of $\mbox{Im}\rho_{12}(t)$ and $\mbox{Re}\rho_{12}(t)$
with the fitting functions given by Eq.~(\ref{ffr12}), where we
vary $\mu-E$ but fixing $d_{L,R}=5 \mu$eV and $\Gamma_{L,R}=25
\mu$eV. The red lines correspond to the sub-exponential fitting
function ($s=1/2$), the blue long-dashed lines are the simple
exponential fitting function ($s=1$), and  the black dashed lines
are the super-exponential fitting function ($s=3/2$).}
\label{fig19}
\end{center}
%\end{center}
\end{figure}

Now we shall extract the relaxation time $T_1$ from
$\mbox{Re}\rho_{12}(t)$ and the decoherence time $T_2$ from
$\mbox{Im}\rho_{12}(t)$ or $\rho_{11}(t)$ using the concept of
half life from the exact numerical solutions and from the fitting
functions (\ref{ffr12}) and (\ref{ffr11}). The results is plotted
in the Figs.~\ref{fig20}-\ref{fig22}.  In Fig.~\ref{fig20} we plot
the relaxation time $T_1$ and the decoherence time $T_2$ from
$\mbox{Re}\rho_{12}(t)$ and $\mbox{Im}\rho_{12}(t)$, respectively,
by varying the chemical potential $\mu-E$, the tunneling rates
$\Gamma_{L,R}$ and the spectral widths $d_{L,R}$ differently.
Fig.~\ref{fig20}(a)-(b) are the plots of $T_1$ and $T_2$ by
varying $\mu-E$ but fixing $(\Gamma_{L,R}, d_{L,R})=(25,5)\mu$eV
(solid lines, corresponding to the non-Markovian processes) and
(5, 25) $\mu$eV (dashed lines for Markovian limit). The results
tell that the decoherence effect (for both $T_1$ and $T_2$) is
large for the small chemical potential $\mu-E$ due to the large
charge leakage effect. Increasing $\mu-E$ reduces the charge
leakage effect, thus also reduces the decoherence effect. When
$\mu-E$ is larger than $\Gamma_{L,R}$ and $d_{L,R}$ [goes up to 30
$\mu$eV in Fig.~\ref{fig20}(a)-(b)], the decoherence effect
quickly reaches to a minimum value (the longest decoherence time
$\sim$ 2 ns). Fig.~\ref{fig20}(c)-(d) plot $T_1$ and $T_2$ by
varying $\Gamma_{L,R}$ but fixing $(\mu-E, d_{L,R})=(10,5)$ (solid
lines, where both the charge leakage effect and the memory effect
are supposed to play an important role) and (30, 25) $\mu$eV
(dashed lines, where both the charge leakage effect and the memory
effect are ignorable). It shows that for a small tunneling rate
between the reservoir and dot ($< \Delta$) the decoherence effect
is weak. Increasing $\Gamma_{L,R}$ enhances the non-Markovian
dynamics effect and also enhances the decoherence (shorting the
relaxation and dephasing times). Fig.~\ref{fig20}(e)-(f) plot
$T_1$ and $T_2$ by varying $d_{L,R}$ but fixing $(\mu-E,
\Gamma_{L,R})=(10,25)$ (solid lines, where the charge leakage
effect is dominated) and (30, 5) $\mu$eV (dashed lines, where the
charge leakage effect is negligible). We find that when the memory
effect must be considered (corresponding to a small spectral
width), the decoherence effect is small (or the relaxation and
dephasing times become longer). Increasing $d_{L,R}$ reduces the
memory or non-Markovian dynamics effect and enhances the
decoherence (shorting the relaxation and dephasing times). It is
interesting to see that in all the cases, the relaxation time is
close to the dephsing time: $T_1 \simeq T_2$ in the Markovian
limit (dashed lines), where for the non-Markovian regime, the
relaxation time is larger than the dephasing time up to twice:
$T_1 \leq 2T_2$ (solid lines).
\begin{figure}[ht]
%\begin{center}
\begin{center}
\includegraphics[width=8.5cm, height=10.5cm]{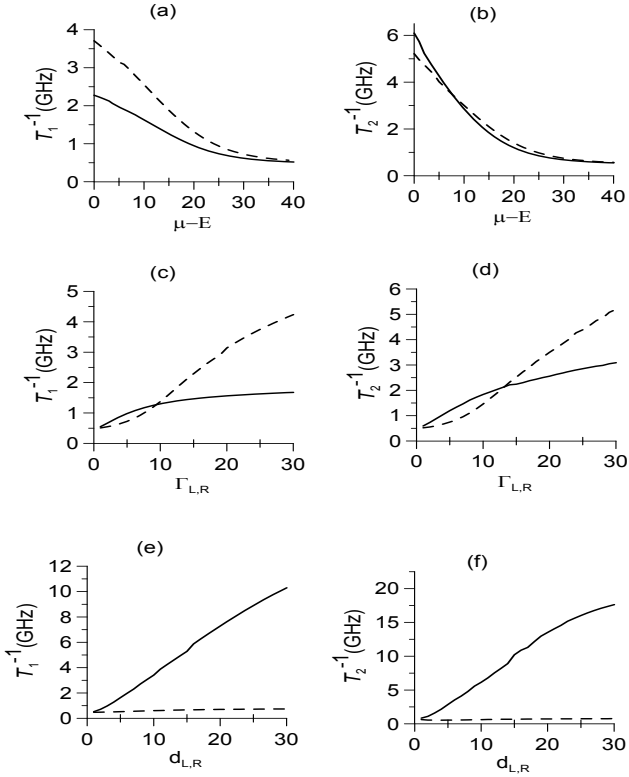}
\caption{The relaxation time $T_1$ and the dephasing time $T_2$
extracted from $\mbox{Re}\rho_{12}(t)$ and
$\mbox{Im}\rho_{12}(t)$, respectively, (a)-(b) by varying the
chemical potential $\mu-E$ but fixing $(\mu-E, d_{L,R})=(10,5)$
(solid lines) and (30, 25) $\mu$eV (dashed lines), (c)-(d) by
varying the tunneling rates $\Gamma_{L,R}$ at $(\mu-E,
d_{L,R})=(10,5)$ (solid lines) and (30, 25) $\mu$eV (dashed
lines), and (e)-(f) by varying the spectral widths $d_{L,R}$ but
fixing $(\mu-E, \Gamma_{L,R})=(10,25)$ (solid lines) and (30, 5)
$\mu$eV (dashed lines). } \label{fig20}
\end{center}
%\end{center}
\end{figure}

As we have pointed out that the decay behaviors for $\rho_{11}$
and $\mbox{Im}\rho_{12}$ are slightly different from the fitting
function in the non-Markovian regime. In Fig.~\ref{fig21}, we
compare the decoherence time $T_2$ extracted from $\rho_{11}(t)$
and $\mbox{Im}\rho_{12}(t)$, respectively, with the same
conditions as used in Fig.~\ref{fig20}, namely by varying the
chemical potential $\mu-E$, the tunneling rates $\Gamma_{L,R}$ and
the spectral widths $d_{L,R}$ differently. As we see although
quantitatively the dephasing time extracted from $\rho_{11}(t)$
and $\mbox{Im}\rho_{12}(t)$ are in the same order, but there are
some obvious differences in certain range of the tunable
parameters where the charge leakage effect play an important role.
This is indeed clearly show in the right-bottom two plots in
Fig.~\ref{fig21} where $\mu-E=30 \mu$eV. When $\Gamma_{L,R}$ are
small ($\leq \Delta/2$) so that the charge leakage is negligible,
the dephasing time $T_2$ extracted from $\rho_{11}(t)$ and
$\mbox{Im}\rho_{12}(t)$ are very close to each other over there.
This again indicates that it is the charge leakage effect that
results in the slightly different decay law for $\rho_{11}(t)$
(follows a sub-exponential decay) and $\mbox{Im}\rho_{12}(t)$ (by
a simple exponential decay) in the non-Markovian regime.
\begin{figure}[ht]
%\begin{center}
\begin{center}
\includegraphics[width=8.5cm, height=10.5cm]{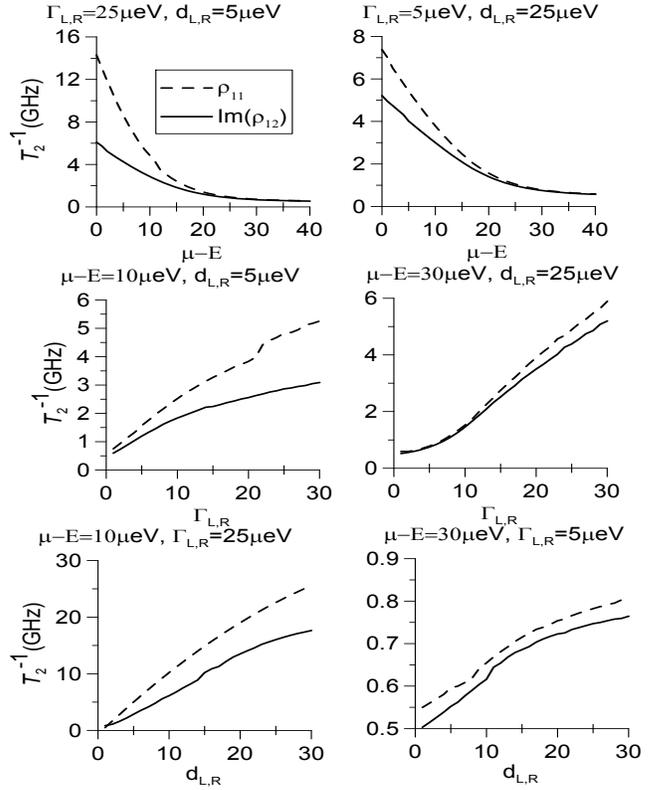}
\caption{The decoherence time $T_2$ extracted from $\rho_{11}(t)$
and $\mbox{Im}\rho_{12}(t)$, respectively, with the same
conditions used in Fig.~\ref{fig20}, namely by varying the
chemical potential $\mu-E$, the tunneling rates $\Gamma_{L,R}$ and
the spectral widths $d_{L,R}$ differently.} \label{fig21}
\end{center}
%\end{center}
\end{figure}

In fact, the original definition of the decoherence for a qubit is
given by the decay of $|\langle+|\rho|-\rangle|$ though it is not
easy to be measured directly in experiments.  Ultimately when the
charge qubit is completely decohered, $\rho_{11}\sim\rho_{22}$ and
$\mbox{Im}(\rho_{12})\sim0$ thus $|\langle+|\rho|-\rangle|\sim0$
at the asymptotic time. We have verified this property in our
exact numerical calculation. Thus the off-diagonal reduced density
matrix element in the energy eigenbasis,
$|\langle+|\rho(t)|-\rangle|$, can be well fitted by $Ae^{-Bt^{s}}
+C$ with $C=0$. In Fig.~\ref{fig22} we compare the results of
$T_2$ extracted from $|\langle+|\rho|-\rangle|$ and
$\mbox{Im}(\rho_{12})(t)$, respectively. It is remarkable that the
dephasing time $T_2$ obtained from $|\langle+|\rho|-\rangle|$ and
$\mbox{Im}(\rho_{12})(t)$ are almost exactly the same in a wide
range of parameters concerned here( $\mu-E$ is from 0 to
$40\mu$eV, $\Gamma_{L,R}$ and $d_{L,R}$ are from 1 to
$30\mu\mbox{eV}$ at $\Delta=10\mu$eV). The decoherence (dephasing)
time $T_2$ obtained here is between 0.2 to 2 ns, except for the
case shown in the left-bottom plot in Fig.~\ref{fig21} where the
decoherence time even smaller.
\begin{figure}[ht]
%\begin{center}
\begin{center}
\includegraphics[width=8.5cm, height=10.5cm]{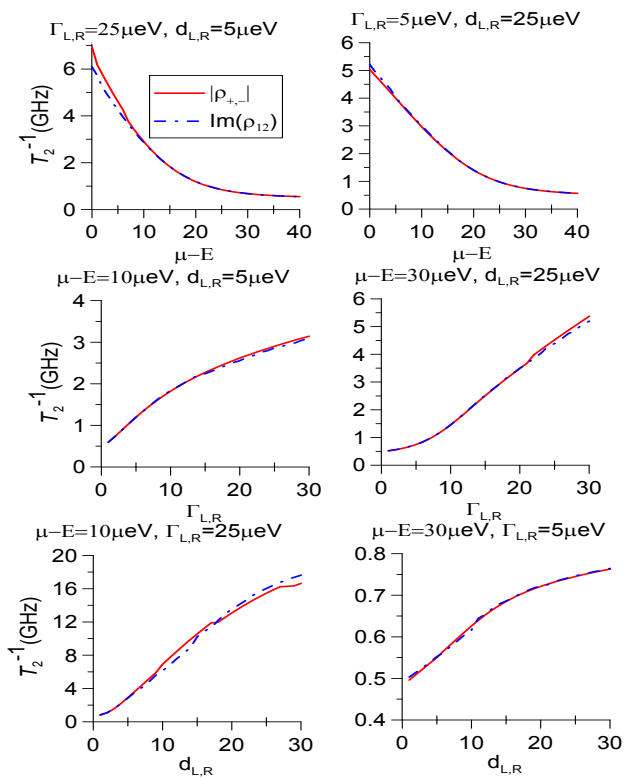}
\caption{The decoherence time $T_2$ extracted from
$|\rho_{+,-}(t)|=|\langle+|\rho|-\rangle|$ and
$\mbox{Im}\rho_{12}(t)$, respectively, with the same conditions
used in Fig.~\ref{fig21}, namely by varying the chemical potential
$\mu-E$, the tunneling rates $\Gamma_{L,R}$ and the spectral
widths $d_{L,R}$ differently.} \label{fig22}
\end{center}
%\end{center}
\end{figure}

Now we shall end this section with a brief summary in the
following. The non-Markovian coherence and decoherence dynamics of
charge qubit is dominated by two major effects, the memory effect
and the leakage effect in the double dot gated by electrode
reservoirs. The former becomes a dominate effect when the time
scale of the reservoirs is comparable to the time scale of the
double dot. The latter becomes an important effect when the
electron tunneling strength between the reservoirs and dots is
tuned to be large. These two characters are suitably described by
the spectral widths and the tunneling rates embedded in the
Lorentzian spectral density (\ref{spectral}) we used.
Strengthening the couplings between the reservoirs and dots and
widening the spectral widths of the reservoirs disturb the charge
coherence in the double dot significantly. However, reasonably
raising up the chemical potentials $\mu_{L,R}$ can suppress charge
leakage and maintain charge coherence. The left uncontrollable
decoherence factor is the spectral width which characters how many
electron states in the reservoirs effectively involving in the
tunneling processes between the dots and the reservoirs. The
smaller the spectral width is (the less the electron states
involve in the electron tunneling), the better the charge
coherence can be maintained. The decay of the charge coherent
oscillation is well described by a simple exponential decay for
the off-diagonal reduced density matrix elements, but the diagonal
ones (populations) are better described by a sub-exponential law
when charge leakage is not negligible. Otherwise simple
exponential decay is better for both the non-Markovian and
Markovian regimes. The relaxation time $T_1$ and the dephasing
time $T_2$ can be extracted from the exact numerical solution
$\mbox{Re}\rho_{12}(t)$ and $\mbox{Im}\rho_{12}(t)$, respectively,
with the result $T_1 \leq 2T_2$ for a broad parameter range we
used.

\section{Conclusion and Discussion}
In this paper, we have developed a non-perturbation theory to
describe decoherence dynamics of electron charges in the double
quantum dot gated by electrodes. We extended the Feynman-Vernon
influence functional theory to fermionic environments and derived
an exact master equation (\ref{emaster}) for the reduced density
matrix of the double dot without including the inter-dot Coulomb
repulsion at beginning. The contributions of quantum and thermal
fluctuations induced by the electron reservoirs are embedded into
the time-dependent transport coefficients (\ref{td-coe}) and
(\ref{ec-ren}) in the master equation. These time-dependent
transport coefficients are completely determined by the
non-perturbation dissipation-fluctuation equations of motion
(\ref{EM-3}). The exact master equation is then further extended
to the double dot in the strong inter-dot Coulomb interacting
regime in terms of Bloch-type rate equations (\ref{i-re}) where
the strong Coulomb repulsion simply leads one to exclude the
states corresponding to a simultaneous occupation of the two dots
from Eq.~(\ref{emaster}). Our theory is developed for a general
spectral density of the reservoirs at arbitrary temperatures and
bias. Other approximated master equations used for the double
quantum dot can be obtained at well defined limits of the present
theory. This non-perturbation decoherence theory allows us to
exploit the quantum decoherence dynamics of the charge qubit
brought up by the tunneling processes between the reservoirs and
dots through qubit manipulations.

We then used the master equation (in terms of the rate equations)
to study the non-Markovian decoherence dynamics of the double dot
charge qubit with the back-action of the reservoirs being fully
taken into account. To make qualitative and also quantitative
understandings of the charge qubit decoherence, we numerically
solve the dissipation-fluctuation integro-differential equations
of motion using a Lorentzian spectral density. We examine the time
dependence of all the transport coefficients from which the time
scales within which non-Markovian processes become important in
the charge coherent dynamics are determined.  The correlation time
of the electron reservoirs (in terms of the spectral widths in the
Lorentzian spectral densities) and the electron tunneling
strengths between the reservoirs and dots (in terms of the
tunneling rates in the Lorentzian spectral densities) characterize
the time scales for the occurrence of non-Markovian processes.
Non-Markovian processes dominate the charge coherent dynamics when
the spectral width is comparable to the inter-dot tunnel coupling
where the memory effect plays an important role, and/or when the
tunneling rates between the reservoirs and dots become strong such
that charge leakage becomes a main effect for decoherence. Raising
up the fermi surfaces of the reservoirs can suppress charge
leakage. The Markovian limit can be reached with a weak tunneling
rate and a large spectral width, where perturbation theory becomes
valid and the spectral density is reduced to a constant. The decay
of the charge coherent oscillation is well described by a simple
exponential law, except for some special regime where the charge
leakage is not negligible such that the evolution of state
populations is better described by a sub-exponential decay. We
also extracted the relaxation time $T_1$ and the decoherence time
$T_2$ consistently from different elements of the reduced density
matrix and obtained a general result $T_1 \leq 2T_2$ which is of
the order of ns or less in a broad parameter range we considered.
These results are ready to be examined in experiments \cite{prs}.

Although we concentrate in this paper the electron charge
coherence (decoherence) dynamics in the double quantum dot system,
the theory we developed in this work can also be applied to
investigate other physical properties, such as quantum transport
phenomena, in various quantum dot structures. In fact, the
spectral density describing the spectral distribution of electron
reservoirs and electron tunneling processes between the reservoirs
and dots has not been well determined experimentally. We used a
Lorentzian spectral density that has been used by others yet still
been waiting for justifying in experiments. The tunneling rates
$\Gamma_{L,R}$ in the Lorentzian spectral density is indeed
tunable in experiment, and have been extracted from current
spectra by assuming a constant spectral density. With the
capability of monitoring the time evolution of electronic
population transfer, we can look at closely the short time
transport properties in this double dot device from which we may
extract the tunneling rates and the spectral widths for the
Lorentzian spectral density or other possible forms of the
spectral density. Otherwise it may be other surroundings (phonons
and fluctuation in impurity configurations, etc.)~that play an
important role in the dynamics of charge qubit decoherence. These
together with other transport properties in various
nanostructures, such as Kondo effect and Fano resonance, etc.,
deserve a separate study. We will leave these research problems to
be addressed in separate papers.

\section*{Acknowledgement}

We would like to thank A. M. van den Brink, S. A. Gurvitz, Y. J.
Yan and X. Q. Li for useful discussions. This work is supported by
the National Science Council of ROC under Contract
No.~NSC-96-2112-M-006-011-MY3.
%%%%%%%%%%%%%%%%%%%%%%%%%%%%%%%%%%%%%%%%%%%%%%%%%%%%%%%%%%%%%%%

\appendix

\section{Derivation of the influence functional}
The propagating function governing the time evolution of the
reduced density matrix is given by Eq.~(\ref{ppg}), in which the
generalized Feynman-Vernon's influence functional is defined by
\begin{align}
\mathcal{F}[\xi^{*},\xi,\eta^{*},\eta] = &\int d\mu(f_{N})
d\mu(f_{0})d\mu(g_{0})\langle f_0|\rho_E(t_0)|g_0\rangle
\nonumber\\ &\times \int\mathcal{D}[f^{*},f,g^{*},g] e^{i
\big\{S_{E}[f^{*},f]-S^*_{E}[g^{*},g]} \nonumber
\\ &~~~~~~~~~~
^{+S_{I}[\xi^{*},\xi,f^{*},f]-S^*_{I}[\eta^{*},\eta,g^{*},g]\big\}},
\end{align}
where $\rho_{E}$ is the initial density operator of the fermion
reservoirs, $f_{0},g_{0},f_{N}$ and their complex conjugates are
Grassmann numbers introduced in the fermion coherent state
representation, $S_{E}$ is the action of the electron reservoirs,
and $S_{I}$ stands for the action of the interaction between the
dots and the reservoirs. Explicitly,
\begin{subequations}
\label{actione}
\begin{align}
iS_{E}&[f^{*},f]= \sum_{l=1,2; k}\Big\{{f^*_{l k}(t)f_{l
k}(t)+f^*_{l k}(t_0)f_{l k}(t_0)\over2} \nonumber\\& +
\int_{t_0}^{t}d\tau \Big[\Big( {\dot{f}^*_{l k}f_{l k}-f^*_{l
k}\dot{f}_{l k}\over2}\Big) -i\varepsilon_{l k}f^*_{l k}f_{l
k}\Big]\Big\}\nonumber\\&~~~~~~~~ =i\sum_{l
k}S_{E,l k}[f^*_{l k},f_{l k}], \\
S_{I}[\xi^{*},& \xi, f^{*},f]= -\sum_{ l k}
\int_{t_0}^{t}d\tau(t_{il k}\xi^{*}_{i} f_{l k} +
t^*_{i l k}f^{*}_{l k}\xi_{i}) \nonumber \\
&~~~~~~~~~~ =\sum_{l k} S_{I,l k}[\xi^{*}_i,\xi_i,f_{l k}^{*},f_{l
k}] .
\end{align}
\end{subequations}

Let the electron reservoirs be initially in a thermal equilibrium
state, then
\begin{align}
\langle f_0|\rho_E(t_0)|g_0\rangle=\prod_{l k}\langle
f_{k0}|{1\over Z}e^{-\beta(\varepsilon_{l k}-\mu_{l})a^{\dag}_{l
k}a_{l k}}|g_{k0}\rangle \label{initialresstae}
\end{align} where $\mu_{L,R}$ are the chemical potentials of the
source and drain electron reservoirs connected to the dots $1$ and
$2$, respectively, and $Z$ is the fermion partition function of
the reservoirs $Z=\prod_{lk}(e^{-\beta(\varepsilon_{l
k}-\mu_{l})}+1)$. Obviously, the influence functional can be
written as $\mathcal{F}[\xi^{*},\xi,\eta^{*},\eta]= \prod_{l
k}\mathcal{F}_{l k}[\xi^{*},\xi,\eta^{*},\eta]$. Furthermore,
since the Hamiltonians are quadratic, the path integrals in the
influence functional can be exactly calculated using either the
Gaussian integrals or the stationary path method. Here we present
the clculation based the stationary path method. The forward
stationary pathes of the electrons in the reservoirs is determined
by
\begin{subequations}
\label{eme}
\begin{align}
& \dot{f}_{l k}(\tau)+ i \varepsilon_{l k}f_{l k}
(\tau)=-it^{*}_{i l k}\xi_{i}(\tau),~\label{emea}\\
& \dot{f}_{l k}^{*}(\tau)-i\varepsilon_{l k}f_{l
k}^{*}(\tau)=it_{i l k}\xi^{*}_{i}(\tau), \label{emeb}
\end{align}
\end{subequations}
with $i=1,2$ for $l=L,R$, respectively. The solutions to the
stationary path equations (\ref{eme}) are
\begin{align}
& f_{l k}(\tau)=f_{l k}(t_{0})e^{-i\varepsilon_{l k}
(\tau-t_{0})}-i t_{i l k}^{*}\int_{t_0}^{\tau}d\tau'
e^{-i\varepsilon_{l k}(\tau-\tau')}\xi_{i}(\tau'),\nonumber \\
& f_{l k}^{*}(\tau)=f_{l k}^{*}(t)e^{i\varepsilon_{l k} (\tau-t)}
-it_{i l k}\int_{\tau}^{t}d\tau'e^{i\varepsilon_{l k}
(\tau-\tau')}\xi^{*}_{i}(\tau'). \label{solfw}
\end{align}
With the similar solutions for the backward stationary pathes, the
$l k$ component of the influence functional is then given by
\begin{widetext}
\begin{align}
\mathcal{F}_{l k} [\xi^*, \xi,\eta^*, \eta]=\exp\Big\{-|t_{il
k}|^{2} \int_{t_0}^{t}d\tau\Big[ \int_{t_0}^{\tau}d\tau'
\Big(e^{i\varepsilon_{l
k}(\tau-\tau')}\eta^{*}_{i}(\tau')\eta_{i}(\tau)
+e^{-i\varepsilon_{l k}(\tau-\tau')}
\xi^{*}_{i}(\tau) \xi_{i}(\tau')\Big) ~~& \nonumber \\
+\int_{t_0}^{t}d\tau'e^{-i\varepsilon_{l k}(\tau-\tau')}
\Big(\eta^{*}_{i}(\tau)\xi_{i}(\tau') -f(\varepsilon_{l
k})[\xi_{i}^*(\tau)
+\eta_{i}^{*}(\tau)][\xi_{i}(\tau')+\eta_{i}(\tau')]\Big)&
\Big]\Big\},
\end{align}
\end{widetext}
where $f(\varepsilon_{l k})$ is the fermi distribution function,
$f(\varepsilon_{l k})={1\over e^{\beta(\varepsilon_{l
k}-\mu_{l})}+1 }$. Sum up contributions from all fermion modes in
the electron reservoirs, we get the influence functional
Eq.~(\ref{inf}) with the dissipation-fluctuation kernels given by
Eq.~(\ref{Kernel}).

\section{Derivation fof the exact propagating function}

In this section we show how to use the solutions of the equations
of motion, (\ref{EM-1}), to determine the time dependent
coefficients in the master equation.  The equation of motion for
$\xi^*_{1,2}$ is just the complex conjugate equation of
$\eta_{1,2}$ while that of $\eta^*_{1,2}$ is the complex conjugate
to the equation of $\xi_{1,2}$. Meantime both
$\xi^{*}_{1,2}(\tau)$ and $\eta_{1,2}(\tau)$ are fixed at $\tau=t$
and both $\eta^{*}_{1,2}(\tau)$ and $\xi_{1,2}(\tau)$ are fixed at
$\tau=t_0$. We only need to solve the set of equations of motion
for $\xi_{1,2}(\tau)$ and $\eta_{1,2}(\tau)$. The solutions to the
equations of motion for $\eta^{*}_{1,2}(\tau)$ and
$\xi^{*}_{1,2}(\tau)$ can be obtained by conjugating the solutions
of $\xi_{1,2}(\tau)$ and $\eta_{1,2}(\tau)$ with corresponding
replacement of the boundary conditions.

Let $\chi(\tau)\equiv \xi(\tau)+\eta(\tau)$, the equations of
motion
 (\ref{EM-1}) becomes
\begin{widetext}
\begin{subequations}
\label{EM-2}
\begin{align}
\dot{\xi}(\tau)+i\begin{pmatrix}E_{1}& T_c \\
T_c & E_{2} \end{pmatrix}\xi(\tau) + \int_{t_0}^{\tau}d\tau' &
\begin{pmatrix}F_{1L}(\tau-\tau')& 0\\0&F_{2R}(\tau-\tau')
\end{pmatrix}
\xi^{}(\tau') \notag \\
& = \int_{t_0}^{t}d\tau'
\begin{pmatrix}F_{1L}^{\beta}(\tau-\tau')&0\\0&F_{2R}^{\beta}(\tau-\tau')\end{pmatrix}
\chi(\tau') \label{EM-2a} \\
\dot{\chi}(\tau)+ i\begin{pmatrix}E_{1}& T_c \\
T_c & E_{2} \end{pmatrix}\chi(\tau) -\int_{\tau}^{t}d\tau' &
\begin{pmatrix}F_{1L}(\tau-\tau')&0\\0&F_{2R}(\tau-\tau')\end{pmatrix}
\chi^{}(\tau')=0 \label{EM-2b}.
\end{align}
\end{subequations}
\end{widetext}
where we have used the following notations for brevity,
\begin{align}
\xi(\tau)=\begin{pmatrix}\xi_{1}(\tau)\\
\xi_{2}(\tau)\end{pmatrix},
~\eta^{}(\tau)=\begin{pmatrix}\eta_{1}(\tau)\\
\eta_{2}(\tau)\end{pmatrix}.
\end{align}
To solve the above equations of motion, we introduce the new
variables $\bar{u}(\tau)$, $u(\tau)$ and $v(\tau)$ such that
\begin{subequations}
\label{transf}
\begin{align}
&\chi(\tau)=\bar{u}(\tau)\chi(t), \label{transf-a}\\
&\xi(\tau)=u(\tau)\xi(t_0)+v(\tau)\chi(t). \label{transf-b}
\end{align}
\end{subequations}
Eq.~(\ref{EM-2}) can then be expressed in terms of  the new
variables $\bar{u}(\tau)$, $u(\tau)$ and $v(\tau)$ as
\begin{widetext}
\begin{subequations}
\label{EM-B}
\begin{align}
\dot{\bar{u}}(\tau)+i\begin{pmatrix}E_{1}& T_c \\
T_c &E_{2}\end{pmatrix}\bar{u}(\tau) - \int_{\tau}^{t}d\tau' &
\begin{pmatrix}F_{1}(\tau-\tau')&0\\0&F_{2}(\tau-\tau')\end{pmatrix}
\bar{u}(\tau')=0 , \label{EM-Ba}\\
\dot{u}(\tau)+i\begin{pmatrix}E_{1}& T_c\\
T_c &E_{2}\end{pmatrix}u(\tau) + \int^{\tau}_{t_0}d\tau' &
\begin{pmatrix}F_{1L}(\tau-\tau')&0\\0&F_{2R}(\tau-\tau')\end{pmatrix}
u(\tau')=0 ,  \label{EM-Bb} \\
\dot{v}(\tau)+i\begin{pmatrix}E_{1}& T_c \\
T_c &E_{2}\end{pmatrix}v(\tau) + \int_{t_0}^{\tau}d\tau' &
\begin{pmatrix}F_{1L}(\tau-\tau')&0\\0&F_{2R}(\tau-\tau')\end{pmatrix}
v^{}(\tau') \notag \\
& =\int_{t_0}^{t} \begin{pmatrix} F_{1L}^{\beta} (\tau-\tau') & 0
\\0&F_{2R}^{\beta}(\tau-\tau')\end{pmatrix} \bar{u}(\tau')
\label{EM-Bc}
\end{align}
\end{subequations}
\end{widetext}
with the boundary conditions $\bar{u}_{ij}(t)=\delta_{ij}$,
$u_{ij}(t_0)=\delta_{ij}$ and $v_{ij}(t_0)=0$, respectively.
Obviously, Eq.~(\ref{EM-Ba}) is the backward version of
Eq.~(\ref{EM-Bb}). Therefore, $\bar{u}(\tau)=u^\dag(t+t_0-\tau)$
for $t_0\le \tau \le t$.

Now $\xi_{1,2}(t)$, $\xi^{*}_{1,2}(t_0)$, $\eta_{1,2}(t_0)$ and
$\eta^{*}_{1,2}(t)$ can be factorized from the the boundary
conditions, $\xi(t_0)=\xi_{0}$, $\eta(t)=\eta_{f}$. Explicitly,
let $\tau=t_0$ for Eq.~(\ref{transf-a}) and $\tau=t$ for
(\ref{transf-b}), we have
\begin{subequations}
\begin{align}
& \xi_{0}+\eta(t_0)=\bar{u}(t_0)
[\xi(t)+\eta_{f}],~ \\
& \xi(t)=u(t) \xi_{0} + v(t) [\xi(t)+\eta_{f}].
\end{align}
\end{subequations}
 Note that $\bar{u}(t_0)=u^\dag(t)$, the above algebraic
equation gives the solution (\ref{SEM-1}). Similarly, let
$\xi^{*}(\tau)=\begin{pmatrix}\xi^{*}_{1}(\tau)&\xi^{*}_{2}(\tau)\end{pmatrix}$
and
$\eta^{*}(\tau)=\begin{pmatrix}\eta^{*}_{1}(\tau)&\eta^{*}_{2}(\tau)\end{pmatrix}$,
we have
\begin{subequations}
\label{SEM-2}
\begin{align} & \xi^{*}(t_0)=
\xi^{*}_{f}\big[I+v^{\dag}(t)(I-v^{\dag}(t))^{-1}\big]u(t)
 \nonumber \\
 &~~~~~~~~~~~~~~~~~~
  -\eta^{*}_{0}\big[I-u^{\dag}(t)(I-v^{\dag}(t))^{-1}u(t)\big], \\
&\eta^{*}(t)=\big[\eta^{*}_{0}u^{\dag}(t)+\xi^{*}_{f}
v^{\dag}(t)\big](I-v^{\dag}(t))^{-1}.
\end{align}
\end{subequations}
Substituting the relations (\ref{SEM-1}) and (\ref{SEM-2}) into
the propagating function (\ref{ppgs}) and using the fact that
$v(\tau)$ is hermitian at $\tau=t$, we obtain the exact
propagating function (\ref{ppg1}) for the double dot gated by bias
electrodes.

\end{document}